\documentstyle[aaspp4]{article}

\def\hatom{\hbox{H}}
\def\hi{\hbox{H\,{\scriptsize I}}}
\def\hii{\hbox{H\,{\scriptsize II}}}

\def\cii{\hbox{C\,{\scriptsize II}}}
\def\cplus{\hbox{C$^+$}}

\def\hh{\hbox{H$_2$}}
\def\hhs{\hbox{H$_2^*$}}

\def\co{\hbox{CO}}
\def\twco{\hbox{CO}}

\def\jone{\hbox{$J = 1\rightarrow 0$}}

\def\vone{\hbox{(1,0) S(1)}}
\def\vtwo{\hbox{(2,1) S(1)}}

\def\vfive{\hbox{(5,3) O(3)}}
\def\vsix{\hbox{(6,4) Q(1)}}

\def\cmv{\hbox{cm$^{-3}$}}
\def\cma{\hbox{cm$^{-2}$}}
\def\kms{\hbox{km\,s$^{-1}$}}
\def\kkms{\hbox{K\,km\,s$^{-1}$}}
\def\ergscmsr{\hbox{ergs\,s$^{-1}$\,cm$^{-2}$\,sr$^{-1}$}}
\def\ergscmsrkms{\hbox{ergs\,s$^{-1}$\,cm$^{-2}$\,sr$^{-1}$\,(km\,s$^{-1}$)$^{-1}$}}
\def\ergscm{\hbox{ergs\,s$^{-1}$\,cm$^{-2}$}}

\def\ev{\hbox{eV}}
\def\av{\hbox{${\rm A_V}$}}
\def\ak{\hbox{${\rm A_K}$}}
\def\texc{\hbox{$T_{exc}$}}
\def\nhh{\hbox{$n$(H$_2$)}}
\def\iuv{\hbox{$I_{UV}$}}
\def\ifir{\hbox{$I_{FIR}$}}
\def\ihh{\hbox{$I_{H_2}$}}
\def\icii{\hbox{$I_{CII}$}}
\def\ico{\hbox{$I_{CO}$}}
\def\xhhs{\hbox{$X_{H_2^*}$}}
\def\xcplus{\hbox{$X_{C^+}$}}

\slugcomment{ To appear in 
    {\it The Astrophysical Journal May 10, 1998 issue, Vol. 498}
}

\lefthead{Pak et al.}
\righthead{Cloud Structure in the Magellanic Clouds}

\begin{document}

\singlespace

\title{
  Molecular Cloud Structure in the Magellanic Clouds: \\
  Effect of Metallicity
}
\author{ Soojong Pak \altaffilmark{1,2}, and 
         D. T. Jaffe \altaffilmark{2} }
\affil{    Department of Astronomy, The University of Texas, 
           Austin, TX 78712, USA }
\author{ Ewine F. van Dishoeck \altaffilmark{3}}
\affil{    Sterrewacht Leiden, P.O.Box 9513, 2300 RA Leiden, 
           the Netherlands }
\and
\author{ L. E. B. Johansson, and R. S. Booth }
\affil{    Onsala Space Observatory, S-439 92 Onsala, Sweden }

\altaffiltext{1}{ Current address: 
    Max-Planck-Institut f\"{u}r extraterrestrische Physik,
    Giessenbachstra\ss e, 85748 Garching, Germany;
    E-mail: soojong@mpe-garching.mpg.de }
\altaffiltext{2}{ Visiting Astronomer, Cerro Tololo Inter-American Observatory, National Optical Astronomy Observatory, which are operated by the Association of Universities for Research in Astronomy, under contract with the National Science Foundation }
\altaffiltext{3}{ The Beatrice M. Tinsley Centennial Visiting Professor, the University of Texas at Austin }

\begin{abstract}

The chemical structure of neutral clouds in low metallicity environments is examined with particular emphasis on the H to \hh\ and C$^+$ to CO
transitions.  
We observed near-IR \hh\ \vone, \vtwo, and \vfive\ lines and the $^{12}$CO \jone\ line from 30 Doradus and N159/N160 in the Large Magellanic Cloud and from DEM S 16, DEM S 37, and LI-SMC 36 in the Small Magellanic Cloud.
We find that the \hh\ emission is UV-excited and that (weak) \twco\ emission always  exists (in our surveyed regions) toward positions where \hh\ and [\cii] emission have been detected.
Using a PDR code and a radiative transfer code, we simulate the emission
of line radiation from spherical clouds and from large planar clouds. 
Because the [\cii] emission and \hh\ emission arise on the surface of the cloud and the lines are optically thin, these lines are not affected by 
changes in the relative sizes of the neutral cloud and the \twco\ bearing 
core, while the optically thick \twco\ emission can be strongly affected.
The  sizes of clouds are estimated by measuring the deviation of CO emission strength from that predicted by a planar cloud model of a given size. 
The average cloud column density and therefore size increases as the metallicity decreases.
Our result agrees with the photoionization regulated star formation theory by McKee (1989).

\end{abstract}


\section{Introduction} 

Stars form in dense, cold molecular clouds.
Measuring the molecular gas content of the clouds is very important if we are to estimate the star formation efficiency and relate it to the properties of the clouds and to their environments.
The total \hh\ mass, however, cannot be measured directly because the lowest levels of \hh\ from which the observable emission can arise have excitation energies (e.g., $\Delta E/k$ $\simeq$ 500 K, $J$ = $2 \rightarrow 0$) too high to be thermally excited in the cold ($< 50$~K) molecular clouds.
In the Milky Way, the $^{12}$CO \jone\ line\footnote{
   The \twco\ \jone\ intensity means the velocity integrated main beam
   brightness temperature, $I = \int T_{MB}\, dv$. }
(hereafter \twco\ \jone ) traces the molecular gas content.
The conversion factor ($X^{GAL}$) between the \hh\ column density and the velocity integrated intensity of CO has been measured via the virial theorem
($X^{GAL}$ = $(2.5-8) \times 10^{20}$ \cma/(\kkms), 
Solomon et al. 1987;
Digel et al. 1997 and references therein), or via gamma-ray emission ($X^{GAL}$ = $1.1-2.8 \times 10^{20}$ \cma/(\kkms), Bloemen et al. 1986;
Digel et al. 1997 and references therein).
The metallicity dependence of the conversion factor has been an issue.
Cohen et al. (1988) and Wilson (1995) used cloud masses determined using the virial theorem to argue that the value of $X$ increases as the metallicity of the individual galaxy decreases.
Arimoto, Sofue, \& Tsujimoto (1996) extend this conclusion to argue that
there are radial increases in $X$ in the Milky Way and M51 corresponding to
radial decreases in metallicity. By contrast, Taylor, Kobulnicky, \& Skillman (1996) showed that some low abundance galaxies have lower $X$, suggesting that a factor other than the abundance (e.g., temperature) can affect the 
measured value of $X$.

Far-UV photons from massive young stars strike the surfaces of nearby molecular clouds\footnote{
In this paper, ``a cloud'' implies an UV-illuminated molecular structure either in isolation or in a complex. }
and produce photodissociation regions or photon-dominated regions (hereafter PDRs, Tielens \& Hollenbach 1985, 1997).
In these surface layers, the far-UV photons dominate the ionization of atoms, the formation and destruction of molecules, and the heating of the gas. 
Inside the PDR, absorption by dust, C, and \hh\ diminishes the far-UV field.
Several authors have constructed PDR models appropriate to conditions in 
the Magellanic clouds, with particular emphasis on the C$^+$/C/CO transition 
(Maloney \& Black 1988\footnote{
Figure 4 in Maloney \& Black (1988) is not correct. The figure should be replaced by Figure 2 in Israel (1988) or by Figure 2 in van Dishoeck \& Black (1988b).}; 
van Dishoeck \& Black 1988b; Lequeux et al. 1994; Maloney \& Wolfire 1997).
In irregular galaxies, where metallicities and dust-to-gas ratios are lower than those in the Galaxy, far-UV photons penetrate deeper into clouds, and dissociate CO molecules to greater depths (Israel et al. 1986).
Therefore, for a cloud with a given column density, the CO column density should be lower at lower metallicity.
If the \twco\ column density is high enough for the \twco\ to self-shield
against photodissociation  ($N(CO)$ $\gtrsim$ $10^{15}$~\cma, van Dishoeck \& Black 1988a), the CO column density will also be high enough for
 the \twco\ \jone\ line to be optically thick, and the \twco\ \jone\ line intensity ($\int T_{MB}\, dv$) will not depend strongly on the metallicity.
In that case, lower \twco\ intensities can only stem from geometrical or
beam-filling effects.
On the other hand, if the cloud column density is not high, most of the CO will be dissociated and the resulting \twco\ line will be optically thin and very weak.
On the surface of the clouds, the destruction and formation of \hh\ molecules are also affected by the change of metallicity, but the mechanism is different from that for CO molecules.
The \hh\ molecules are dissociated by far-UV photons attenuated by dust or 
by \hh\ self-shielding.
If \hh\ self-shielding dominates over dust attenuation, the \hh\ destruction rate is independent of the dust abundance.
On the other hand, the \hh\ formation rate is proportional to the dust abundance, because \hh\ reforms on the surfaces of dust grains.

The Magellanic Clouds are the best targets to test PDR models that include metallicity effects because of their proximity ($d^{LMC}$ = 50.1~kpc and $d^{SMC}$ = 60.3~kpc, Westerlund 1990), their low metal abundance ($Z_C^{LMC}$ = 0.28, $Z_O^{LMC}$ = 0.54, $Z_C^{SMC}$ = 0.050, and $Z_O^{SMC}$ = 0.21, where $Z$ is normalized to the Galactic value; Dufour 1984), and their low dust-to-gas ratio ($\rho_{dust}^{LMC}$ = 0.25 and $\rho_{dust}^{SMC}$ = 0.059, where $\rho$ is normalized to the Galactic value; Koornneef 1984).
In this paper, we observed the Magellanic Clouds in the near-IR \hh\ emission lines and in the \twco\ \jone\ line (see Sections \ref{sec:obs5} and \ref{sec:results}).
We compare the line intensities of \hh\ \vone, \twco\ \jone, and [\cii] 158 \micron\ emission from the PDRs in the Magellanic Clouds with those from Galactic star formation regions (see Section~\ref{sec:compare}).
Section~\ref{sec:models} discusses the numerical PDR models which we compare to the observed data to learn how metallicity changes affect the chemical structure of the Galactic clouds and the clouds in the Magellanic Clouds.

\section{Observations}   \label{sec:obs5}

Some limited regions in the Magellanic Clouds were previously observed in the \hh\ lines (Koornneef \& Israel 1985; Israel \& Koornneef 1988; Kawara, Nishida, \& Taniguchi 1988; Israel \& Koornneef 1992; Krabbe et al. 1991; Poglitsch et al. 1995).
However, the published [\cii] and \twco\ data (Johansson et at. 1994; Poglitsch et al. 1995; Israel et al. 1996) cover more extended regions than the existing \hh\ maps.
We observed near-IR \hh\ emission lines from the Magellanic Clouds with the University of Texas Near-IR Fabry-Perot Spectrometer whose equivalent-disk size\footnote{
    The equivalent-disk size, $\theta_{ED}$, is the diameter of 
    a cylindrical  beam whose solid angle is same 
    as the integrated solid angle of the actual beam pattern.}
 ($\theta_{ED}$ = 81\arcsec) is comparable to those of the existing [\cii] data ($\theta_{ED}$ = 68\arcsec) and \twco\ \jone\ data ($\theta_{ED}$ = 54\arcsec).
We also observed \twco\ \jone\ emission at positions where no emission had been detected at the sensitivity of the existing \twco\ surveys.

\subsection{ Near-IR \hh\ Emission Lines }

We observed the \hh\ \vone\ and \vtwo\ lines in 1994 December, and \vone\ and \vfive\ lines in 1995 October, at the Cerro Tololo Inter-American Observatory
1.5 m telescope using the University of Texas Near-Infrared Fabry-Perot Spectrometer (UT~FPS, Luhman et al. 1995).
The instrument was designed to observe extended, low surface brightness line emission, and has a single 1~mm diameter InSb detector to maximize the area-solid angle product. 
This product, $A\Omega$, depends on the telescope coupling optics but can be made as large as $4.5 \times 10^{-3}$ cm$^2$~sr.
The beam from the $f/30$ secondary of the telescope is collimated and guided to an H-band or K-band Fabry-Perot interferometer and to a $\sim 55$~K (solid nitrogen temperature) dewar in which a doublet camera lens (focal length = 20~mm) focuses onto the detector.
A 0.5~\% (for the \hh\ \vfive\ line) or a 1~\% (for the \hh\ \vone\ and \vtwo\ lines) interference filter in the dewar selects a single order from the Fabry-Perot interferometer.
We used a collimator of focal length = 838~mm in 1994 December run and a collimator of focal length = 686~mm in 1995 October.
The change of the collimator affected the beam size and spectral resolution (Table~\ref{tbl-1}).
Since the coupling of the 686 mm collimator with the telescope optics is better, the beam profile in 1995 October is much closer to a box-function than the 1994 December beam profile (Figure~\ref{fig-01}).

An automatic alignment routine aligned the Fabry-Perot etalon by executing every $5-8$ minutes (Luhman et al. 1995).
The Fabry-Perot etalon maintained alignment for $15-30$ minutes, but the ambient temperature changes caused the plate separation to drift by the equivalent of $2-5$~\kms\ per minute.
Using telluric OH lines (Oliva and Origlia 1992), we calibrated the wavelength scale to within $\pm 30~\kms$.

We operated the Fabry-Perot interferometer in {\it scanning mode} at selected positions and in {\it frequency switching mode} at most positions.
In the scanning mode, the plate separation of the etalon was varied to cover $\pm 200$~\kms\ centered at the \hh\ line in 15 sequential steps.
Figure~\ref{fig-02} shows observed \hh\ \vone\ and \vfive\ lines at the 30 Doradus (0, 0) position (see the object list in Table~\ref{tbl-3}), and telluric OH lines.
The OH (9,7)~R$_2$(2) ($\lambda$ = 2.12267~\micron, wavelength in air, Oliva \& Origlia 1992) and 
the (9,7)~R$_1$(1) ($\lambda$ = 2.12440~\micron) lines are within the \hh\ \vone\ scan range.
Figure~\ref{fig-02} shows the red-wing of the OH
(4,2)~P$_1$(3) ($\lambda$ = 1.61242~\micron) line, and the OH (4,2)~P$_1$(2) ($\lambda$ = 1.60264~\micron) line which was displaced by one free spectral range and penetrated through the blue side of the order sorting filter at $V_{LSR}$ $\simeq$ 420~\kms .
The typical intensity of the OH (4,2)~P$_1$(3) and (4,2)~P$_1$(2) lines is $\sim 4 \times 10^{-3}$ \ergscmsr\ which is more than $10^3$ times the \hh\ \vfive\ intensity.

The OH intensity fluctuates spatially and temporally.
The $1/f$ power spectrum of the temporal fluctuations limits the sensitivity of the system, e.g., in K-band, the OH noise becomes important at 30 seconds (coadded integration time) per step when the switching interval between the source position and an off-source ``sky'' position is two minutes (Luhman et al. 1995).
In the observations of the Magellanic Clouds, we chopped the secondary mirror from the source to $\Delta \alpha$ = $+16\arcmin$ or $-16\arcmin$ at 0.25~Hz in H-band and 0.5~Hz in K-band.
Each spectral step (at the same etalon plate separation) has one chopping cycle which consists of four exposures: $object$ $\rightarrow$ $sky$ $\rightarrow$ $sky$ $\rightarrow$ $object$.
Each exposure had an integration time of 0.5 or 1 second.

In the frequency switching mode, we tuned the Fabry-Perot interferometer to the line wavelength, $\lambda_{on}$, and a nearby wavelength free of line emission, $\lambda_{off}$. 
One observing cycle consists of four steps: $\lambda_{on}$ $\rightarrow$ $\lambda_{off}$ $\rightarrow$ $\lambda_{off}$ $\rightarrow$ $\lambda_{on}$.
As is the case in the scanning mode, each step consists of one chopping cycle.
We tried to place $\lambda_{off}$ away from wings of the \hh\ instrument spectral profile for $\lambda_{on}$ and at positions away from telluric OH lines (see Table~\ref{tbl-2}).

For flux calibration, we measured HR~1713 (B8I, $m_K$ = $+0.18$~mag) 
in 1994 December and HR~8728 (A3V, $m_H$ = $+1.03$~mag and $m_K$ = $+1.00$~mag) in 1995 October.
Even though the beam sizes of the 1994 December run and the 1995 October run are different from each other, \hh\ \vone\ intensities measured at the same positions on both runs agree to within the errors.
The absolute flux calibration is accurate to $\pm 20$~\%.

\subsection{ \twco\ $J=1-0$ Emission Line } \label{sec:obs-co}

We observed \twco\ \jone\ line in 1995 December at the SEST\footnote{ 
  The Swedish-ESO Submillimeter Telescope, SEST, is operated jointly by 
  ESO and the Swedish National Facility for Radio Astronomy, 
  Onsala Space Observatory at Chalmers University of Technology.}
located on La Silla in Chile.
The beam size (FWHM) of the SEST is 45\arcsec.
The \twco\ intensities presented in this paper are the main beam brightness temperatures, $T_{MB}$ (see the convention by Mangum 1993).
The online temperature, $T^{*}_{A}$, has been converted to $T_{MB}$ by dividing by 0.7 (which is the product of the forward spillover and scattering efficiency, and the main beam efficiency).
We used frequency switching to gain a factor of 2 in observing time and to avoid possible emission from reference positions. 
Since this method can leave residual ripples in the spectra, the detection limit for weak signals is determined by the ripple.
For some positions, we complemented the frequency switched data with beam switched data with a reference beam about 12\arcmin\ away in azimuth.
For example, in the 30 Doradus (0, $-6\arcmin$) region, it turned out to be impossible to use the frequency switched data.

We mapped \twco\ \jone\ spectra around positions where we detected \hh\ emission and where the CO emission was below the lowest contour level (3 \kkms) of previously published CO maps (Booth 1993; Johansson et al. 1994). 
The CO maps are fully sampled on 20\arcsec\ grids within each 81\arcsec\ \hh\ beam.
In a single 45\arcsec\ beam, the typical root-mean-square noise is 0.07 K for a channel spacing of 0.45 \kms, and
the typical $1\sigma$ statistical uncertainty of the intensity  $I$ = $\int T_{MB}\, dv$,  
integrated over a velocity range of 15 \kms, is 0.2~\kkms.

\section{Results}   \label{sec:results}

In low metallicity objects, the low dust abundance allows far-UV photons to penetrate for long distances beyond the \hii/H boundary.
One possible scenario is that the transparency to UV photons leads to substantial regions of neutral gas where far-UV photons have dissociated CO but where self shielding permits hydrogen to be primarily in the form of \hh.
Especially in the N159/N160 complex, the published maps show the [\cii] 158 \micron\ emission (Israel et al. 1996) extending farther than the \twco\ \jone\ emission (Johansson et al. 1994).
These existing observations make such a possibility seem reasonable in the LMC.

Near-IR \hh\ emission in response to far-UV radiation provides a direct test for the presence of molecular gas near cloud boundaries, albeit with no information about column density or total abundance.
We selected observing targets in regions of the LMC where high spatial resolution ($\sim 1\arcmin$) [\cii] maps and \twco\ \jone\ maps existed (Booth 1993; Johansson et al. 1994; Poglitsch et al. 1995; Israel et al. 1996).
In the SMC, our \hh\ observations cover positions for which there were \twco\ \jone\ data available (Rubio et al. 1993) but do not coincide with the observed [\cii] positions (Israel \& Maloney 1993).
Table~\ref{tbl-3} lists the observed sources and their reference ($0, 0$) positions.

\subsection{ \hh, \twco, [\cii], and far-IR } \label{sec:far-IR}

The \hh\ \vone\ line intensity in the surveyed regions is $<$ $4 \times 10^{-6}$ \ergscmsr, except in the central regions of 30 Doradus and N160.
About 70\% of the observed points have detections of the \vone\ line with a significance of $2\sigma$ or more.
The \hh\ \vone\ intensities are listed in Table~\ref{tbl-4}, and the \hh\ \vtwo\ and \vfive\ intensities in Table~\ref{tbl-5}.
Reddening toward the stars in the LMC has been measured using (\ub) and (\bv) colors.
The Galactic foreground extinction, \av, toward the LMC is $0.23 \pm 0.07$ mag (Greve, van Genderen, \& Laval 1990; Lee 1991).
Applying \ak\ = 0.112~\av (Rieke \& Lebofsky 1985), the foreground \av\ implies a negligible \ak\ of $\sim 0.03$ mag.
The molecular clouds in the LMC may, however, extinguish the \hh\ emission from their own back sides.
Section \ref{sec:without_epsilon} discusses the \hh\ emission from the back side of the clouds.

Table~\ref{tbl-4} lists the \twco\ \jone\ and [\cii] intensities convolved to the beam size of the \hh\ observations ($\theta_{ED}$ = 81\arcsec).
We assumed that the beam profile of the University of Texas Fabry-Perot Spectrometer is a box function.

The columns for far-IR intensity and dust temperature in Table~\ref{tbl-4} are from the IRAS data processed at IPAC\footnote{
  IPAC is funded by NASA as part of the IRAS extended mission program 
  under contract to JPL.}
using standard HIRES processing. 
The HIRES processor enhances the spatial resolution of IRAS images using 
a Maximum Correlation Method algorithm (Aumann, Fowler, \& Melnyk 1990).
The effective resolutions are $1\farcm0 - 1\farcm6$ at 60 \micron\ and $1\farcm6 - 2\farcm4$ at 100 \micron .
We calculate the integrated far-IR intensity using the approximation of Helou et al. (1988):
\begin{equation}  \label{eq:I_FIR}
  \ifir = 1.26 \times 10^{-5}\ (2.58\ I_{60\mu m} + I_{100\mu m})\ ,
\end{equation}
where $I_{60\mu m}$ and $I_{100\mu m}$ are in units of MJy/sr and \ifir\ is in \ergscmsr.
This equation is valid within 20 \% errors, when $31$ $<$ $T_{dust}$ $<$ 58 K.
We deduce the dust temperature, $T_{dust}$, from the ratio of 60 \micron\ to 100 \micron\ intensity:
\begin{equation} \label{eq:T_dust}
  \frac{I_{60\mu m}}{I_{100\mu m}} 
  = \left( \frac{\nu_{60\mu m}}{\nu_{100\mu m}} \right)^4
    \frac{\exp(h \nu_{100\mu m} / k T_{dust}) - 1}
         {\exp(h \nu_{60\mu m}  / k T_{dust}) - 1}\ ,
\end{equation}
where $h$ is the Planck's constant, $k$ is the Boltzmann's constant, and $\nu$ is the frequency in units of Hz.
In the above equation, we assumed that the 60 \micron\ emission is not dominated by small ($D \simeq 5$ nm), thermally-spiking, grains but by large grains in steady state temperatures (Draine \& Lee 1984), and that the dust emissivity ($Q_\nu$) is proportional to $\nu^n$ where the dust emissivity index, $n$, is 1.

\subsection{ H$_2$ Results }

\subsubsection{ SMC }

In the SMC, we detected \hh\ \vone\ emission, with $S/N > 2$, near IRAS sources in the DEM S 16, DEM S 37, and LI-SMC 36 regions.
Toward supernova remnants, e.g., 
DEM S 37 ($+0\farcm38, -3\farcm35$) and LI-SMC 36 ($-3\arcmin, 0$), we did not detect \hh\ emission at a $1\sigma$ level of  $\sim 2 \times 10^{-6}$ \ergscmsr.

\subsubsection{ LMC: 30 Doradus } \label{sec:LMC}

Figure \ref{fig-03} shows the far-IR, \twco\ \jone, and \hh\ \vone\ intensity maps of the 30 Doradus region.
At the 30 Doradus ($0, 0$) point, the intensity of the \hh\ \vone\ line, \ihh, is $1.1 \times 10^{-5}$ \ergscmsr\ in our 81\arcsec\ beam (hereafter, \ihh\ denotes the intensity of the \hh\ \vone\ line).
Poglitsch et al. (1995), using their imaging NIR-spectrometer FAST, observed intense \hh\ \vone\ emission around the central cluster R136 and showed that the \hh\ source appears highly fragmented ($< 5\arcsec$ or 1 pc scale) with a typical intensity of $\sim 1.6 \times 10^{-4}$ \ergscmsr.

Our observations show that the \hh\ emission in 30 Doradus is very extended, $> 40$ pc, compared to the known extent of the CO J=1$\rightarrow$0 emission ($\sim 35$ pc, see the right map in Figure \ref{fig-03}).
At 1\farcm5 (or 22 pc) from the ($0, 0$) position, the \hh\ \vone\ intensity is only a factor of two lower than at the peak (see Figure~\ref{fig-03}).
We also detected faint \hh\ \vone\ emission at ($0, -6\arcmin$), \ihh\ = $1.6 \times 10^{-6}$ \ergscmsr, where \twco\ \jone\ emission had not been detected during the survey of the ESO-SEST Key Program (Booth 1993).

\subsubsection{ LMC: N159/160 }

The N159/N160 \hii\ complexes are 40\arcmin\ south of 30 Doradus.
The [\cii] line (Israel et al. 1996) and the far-IR continuum distributions (left map in Figure \ref{fig-04}) show that the far-UV fields are strong both near N159 and N160.
On the other hand, the CO intensity around N159 is more than four times stronger than that around N160.

The H$_2$ \vone\ line had been detected at ``N159 Blob'' (a compact \hii\ source with a size of $8\arcsec \times 6\arcsec$, Heydari-Malayeri \& Testor 1985) by several groups.
In Table \ref{tbl-6}, we compare the previous data with our new results.
The flux increases as the beam size increases. 
This suggests that, if there is a single source, the emission region is more extended than $\sim 20\arcsec$.
Alternatively, there could be clumpy emission filling our $81\arcsec$ beam
with an area covering factor $\sim20$\% of the covering factor in the
inner $6\arcsec \times 6\arcsec$ region.
In our \hh\ survey, we observed very extended ($> 5\arcmin$ or 70 pc) \hh\ \vone\ emission from the N159/N160 \hii\ complex (see the right map in Figure~\ref{fig-04}).
In the N159 region, we detected \hh\ emission where the CO cloud complex is bright and extended (see Figure 1 in Johansson et al. 1994).
In the N160 region, however, we also detected \hh\ emission beyond the lowest \twco\ \jone\ contour level (3 \kkms) in the map by Johansson et al. (1994)\footnote{
  See also Figure 2 in Israel et al. (1996) which used the data of Johansson et al. (1994).
  More recently, the SEST Key Programme has observed more positions in 30 Doradus and N159/N160 (Johansson et al. 1998). }.
In spite of the weak or absent CO emission in the N160 region, the \hh\ observations indicate that the size of the molecular cloud complex is as big as that in the N159 region.

\subsection{ \twco\ \jone }

At several positions in the outer regions of 30 Doradus and N160, we detected \hh\ emission where earlier CO surveys failed to detect the \jone\ line.
In order to determine if all CO is dissociated at these positions, we observed the regions again in the \twco\ \jone\ line (see Section \ref{sec:obs-co}) with higher sensitivity ($\sigma \simeq 0.2$ \kkms) than the previous observations of which the lowest contour level was 3~\kkms\ (Booth 1993; Johansson et al. 1994).

We made a fully sampled \twco\ map inside the UT FPS beam at ($0, -6\arcmin$) in 30 Doradus, and detected \twco\ \jone\ at a level of 0.5~\kkms\ (see Figure \ref{fig-03}).
We also made \twco\ maps in the outer regions of N160 where the previous \twco\ \jone\ map did not show any \twco\ emission, and sampled some positions in N159 to confirm the flux calibration of the new observations (see Figures \ref{fig-04} and \ref{fig-05}).
We plot the contours at logarithmic intervals to emphasize the edges of the molecular cloud complexes; note the dense contour lines on the northwestern side and on the western side of the N160 complex.
The \twco\ \jone\ emission regions cover the \hh\ emission regions except the ($-2\farcm5, +7\arcmin$) and the ($-2\arcmin, +10\arcmin$) positions where the CO cloud complex fills $50-70$ \% of the \hh\ beams.

\subsection{ H$_2$ Excitation Mechanism } \label{sec:excitation}

Table~\ref{tbl-5} lists the observed \hh\ \vtwo/\vone\ and \vfive/\vone\ line ratios in the Magellanic Clouds.
Observations of the \hh\ lines in archetypal shocked regions, e.g., Orion BN-KL, HH 7, and the supernova remnant IC 443, show that the \hh\ ratios of \vtwo/\vone\ are almost constant $\sim 0.08$ or $\texc \simeq 2 \times 10^3$ K (Burton et al. 1989; Richter, Graham, \& Wright 1995).
Assuming that the excited levels are in LTE and $\texc = 2000$~K, the \vfive/\vone\ ratio should be only $\sim 9 \times 10^{-5}$.

Molecular hydrogen in PDRs absorbs $91.2-110.8$~nm photons in the 
$B^1\Sigma^+_u - X^1\Sigma^+_g$ Lyman and 
$C^1\Pi_u - X^1\Sigma^+_g$ Werner bands.
About 15\% of the electronically excited molecules are dissociated (Draine \& Bertoldi 1996). 
The remaining 85\% of the excitations result in populations of various ro-vibration levels of the ground electronic state.  
If \nhh\ $<$ $5 \times 10^4\ \cmv$, the relative line intensities arising in UV-excited \hh\ are insensitive to density or to UV field strength (Black \& van Dishoeck 1987).
In this pure-fluorescent transition case, the \hh\ ratio of \vtwo/\vone\ is 0.56 and \vfive/\vone\ is 0.38.
At densities $\geq$ $5 \times 10^4\ \cmv$ (Luhman et al. 1997), the collisional de-excitation of UV-pumped \hh\ begins to affect the ro-vibrational level populations.
If the PDR boundaries become sufficiently warm ($T > 1000$ K) and dense ($\geq$ $5 \times 10^4\ \cmv$), collisional excitation thermalizes the low vibrational ($v \leq 2$) level populations (Sternberg \& Dalgarno 1989; Luhman et al. 1997). 

The detections of \hh\ \vfive\ line in the Magellanic Clouds (Table~\ref{tbl-5}) verify the UV-excitation of \hh.
The observed \hh\ line ratios from the N160 region are those expected for pure-fluorescence.
The ratios toward other regions show that the \hh\ ro-vibration levels may be affected somewhat by collisions. 
In 30 Doradus, the peak \hh\ \vone\ intensity is $2.3 \times 10^{-4}$ \ergscmsr\ (Poglitsch et al. 1995) which is brighter than the maximum predicted by our PDR models ($\ihh$ $\leq$ $9.6 \times 10^{-5}$ \ergscmsr, see Figure~\ref{fig-12}).
The models may underestimate the intensity by as much as a factor of two 
in regions with high densities and high UV fields by neglecting the large
increase in UV pumping in the very warm (420-600 K) region at the cloud
edge.  Alternatively, there may be some collisional excitation of \hh\ 
v=1 as well as collisional de-excitation of the high v states, as one finds in clouds subjected to intense far-UV fields (Luhman et al. 1997).

\section{ Comparing with Galactic Clouds } \label{sec:compare}

We compare the \hh\ \vone, [\cii] 158 \micron, \twco\ \jone, and far-IR emission from the clouds in the Magellanic Clouds (see Table~\ref{tbl-4}) to the emission from star forming clouds in Orion and NGC 2024 for which we have comparable data sets (see Table~\ref{tbl-7}).
In the LMC, we select positions where complete \hh, [\cii], and \twco\ data sets exist: 5 positions in the 30 Doradus region, 4 positions in the N159 region, and 8 positions in the N160 region.
In the SMC, we use 4 positions with only \twco\ and \hh\ data.

\subsection{ Data from NGC 2024 and Orion A Star Formation Regions }
    \label{sec:galactic_data}

For the Galactic cloud data, we make use of the published \hh\ \vone, [\cii] 158 \micron\, and \twco\ \jone\ data in Schloerb \& Loren (1982), Stacey et al. (1993), Jaffe et al. (1994), Luhman et al. (1994), Luhman \& Jaffe (1996), and Luhman et al. (1997).
The far-IR data are from the HIRES processed IRAS data (see Section~\ref{sec:far-IR}).
In the Orion molecular cloud, Stacey et al. (1993) made two [\cii] strip maps in Right Ascension, both of which pass close to the \twco\ \jone\ peak.
The [\cii] flux along the strips was ``bootstrapped'', integrated by assuming zero flux at the ends of the cut and summing the chopped differences.
The data from {\it cut 1} (observed west to east) and {\it cut 2} (observed east to west) are not in complete agreement; therefore, we take only those data with:
\begin{equation}
  2 \frac{\mid I_{cut1} - I_{cut2} \mid}{I_{cut1} + I_{cut2}} < 0.4\ .
\end{equation}
We also exclude positions where the IRAS data are saturated, i.e., $I_{60\mu m}$ $>$ 1500~MJy/sr, and where the \hh\ ro-vibrational level populations begin to show effects of collisional de-excitation (\vsix/\vone\ $<$ 0.15, see Table~2 in Luhman \& Jaffe 1996).
Some positions where collisional de-excitation was unlikely were not observed in the \hh\ \vsix\ line but are included in the data we use (see the discussion in Section \ref{sec:hh_cii}).
Table~\ref{tbl-7} lists the compiled data sets in the Galactic Clouds: 4 positions from the Orion A molecular cloud (hereafter ``Orion''), and 11 positions from the cloud associated with the NGC 2024 \hii\ region (hereafter ``NGC 2024'').

Figure \ref{fig-06} compares the data from 15 positions in the Galactic clouds with the data by Jaffe et al. (1994) who presented \twco\ \jone\ and [\cii] data from NGC 2024, and showed that the different parts of the source have very different distributions on a $\icii$ versus $\ico$ plot.
The left plot in Figure \ref{fig-06} shows the distinction between 
{\it the cloud proper zone} ($\Delta \alpha$ $>$ $-10\arcmin$ with respect to NGC 2024 IRS 1, shown as open circles) and 
{\it the western edge zone} ($\Delta \alpha$ $<$ $-10\arcmin$, shown as open triangles) in NGC 2024.
Data from the cloud proper zone agree with the PDR models of Wolfire et al. (1989), while many of the \icii/\ico\ ratios toward the western edge zone are much higher than any of the model ratios. 
Jaffe et al. (1994) interpreted the western edge zone results as implying that the mean column density of clouds decreases to the west.
The right plot in Figure \ref{fig-06} shows the location in the $\icii-\ico$ space of the data sets in Table~\ref{tbl-7}, which will be used to compare with those from the Magellanic Clouds.

\subsection{ Far-IR Data } \label{sec:beam}

\subsubsection{ \iuv\ versus \ifir } \label{sec:I_UV}

The intensity of the interstellar far-UV radiation field is usually expressed in terms of the scaling factor \iuv, the mean intensity in the solar neighborhood (Draine 1978; also see footnote 2 in Black \& van Dishoeck 1987).
The critical part of the UV range for \hh\ fluorescence, C ionization, and CO dissociation is 91.2 $< \lambda <$ 113~nm (Black \& van Dishoeck 1987; van Dishoeck \& Black 1988a).
The intensity of the $\iuv = 1$ field integrated over 91.2 $< \lambda <$ 113~nm is $3.71 \times 10^{-5}$ \ergscmsr.
Note that some authors (e.g., Tielens \& Hollenbach 1985) use Habing's (1968) determination of the local far-UV field, $1.3 \times 10^{-4}$ \ergscmsr, for the integrated intensity between 6 $< h\nu <$ 13.6~\ev\ or 91.2 $< \lambda <$ 206.6~nm, and use the symbol $G_\circ$ to indicate the degree of enhancement over this standard integrated intensity.
The corresponding intensity of the Draine field integrated over the same band as used for $G_\circ$ is $2.13 \times 10^{-4}$ \ergscmsr\ (for a detailed comparison between the Draine field and the Habing field, see Draine \& Bertoldi 1996).

Most of the far-UV energy is absorbed by grains and reradiated in the far-IR:
\begin{equation} \label{eq:G0-I_FIR}
  \ifir = 2 \times ( 2.13 \times 10^{-4} )\ \iuv\ \ergscmsr, 
\end{equation}
where the factor of 2 accounts for incident radiation at longer wavelengths, $I_\nu(\lambda > 206.6$~nm) (Wolfire, Hollenbach, \& Tielens 1989).

\subsubsection{ \ifir\ versus $T_{dust}$ }

We can use the far-IR emission to normalize the various line intensities to compensate for beam filling factor effects.
Before we do so, however, we need to understand how the far-IR emission arises.
Figure~\ref{fig-07} shows  \ifir\ versus $T_{dust}$ for the Galactic star forming cloud positions and the Magellanic Clouds positions in our sample.
We derive \ifir\ from the measured 60 \micron\ and 100 \micron\ intensities using Equation \ref{eq:I_FIR}, and derive $T_{dust}$ using Equation \ref{eq:T_dust}.
Assuming that the dust is heated by an external far-UV field, we can approximate the results of PDR models:
\begin{equation}
  T_{dust} = 13.5\, 
    \left( \iuv\, \frac{T_{eff}}{3 \times 10^4} \right)^{1/5}\ {\rm K}\, 
\end{equation}
where $T_{eff}$ is the equivalent stellar surface temperature which would produce the incident UV field (Hollenbach, Takahashi, \& Tielens 1991; Spaans et al. 1994). 
Assuming $T_{eff}$ = $3 \times 10^4$ K, the expected relation between $T_{dust}$ and \ifir\ is:
\begin{equation} \label{eq:tfir_model}
  \log \ifir = -9.02 + 5 \log T_{dust},
\end{equation}
which is shown as a dotted line in Figure \ref{fig-07}.

The observed $T_{dust}$ and \ifir\ distributions in Orion and NGC 2024 agree with the model in Equation~\ref{eq:tfir_model}.
\ifir\ is proportional to the beam filling factor, while $T_{dust}$, which was deduced from the ratio of $I_{60\micron}/I_{100\micron}$, is independent of the beam filling factor.
With a beam size of $\sim 1\arcmin$ (or $\sim 0.12$ pc at the distance of Orion; 415 pc, Anthony-Twarog 1982), the projected beam filling factor for the clouds in Orion and NGC 2024 is $\sim 1$, which explains the agreement between the observed $T_{dust}-\ifir$ relation and the model.
If we assume that the dust size distribution is independent of metallicity and that the clouds are optically thick in the far-UV, the $T_{dust}$-$\ifir$ relation should be independent of the dust abundance (see Section \ref{sec:self-shielding} for more discussion).
For the Magellanic Clouds, the observed \ifir\ values are an order of magnitude weaker than the values predicted by the $T_{dust}-\ifir$ relation.
This difference implies a beam filling factor in the Magellanic Clouds of $\sim 0.1$ ($d^{LMC}=50.1$ kpc, Westerlund 1990). 

\subsection{ Line Intensities Divided by \ifir } \label{sec:dividefir}

The line and continuum emission we describe here arises in the layers of molecular clouds where UV photons influence the chemistry and the physical conditions.
Various parameters affect the emergent intensity: 
\begin{eqnarray} 
  I_{i} & = & \eta\, \epsilon_{i}\, f_{i}(n_H, \iuv, \delta v_D, Z),
            \label{eq:intensities} \\
  \ifir & = & \eta\, \epsilon_{FIR}\, 4.26 \times 10^{-4}\, \iuv\ , 
            \nonumber
\end{eqnarray}
where $I_{i}$ is the {\it observed} \twco\ \jone, [\cii] 158 \micron, or \hh\ \vone\ emission line intensity, and \ifir\ is the {\it observed} far-IR continuum intensity in units of \ergscmsr.
$f_i$ is the combined intensity arising from the front and back sides of a plane-parallel model cloud which fills the beam.
The front and back surfaces of the cloud are exposed to external far-UV radiation, and are perpendicular to the line-of-sight. 
$f_{i}$ depends on the hydrogen number density, $n_H$ = $n(\hatom)$ + $2 n(\hh)$, where $n(\hatom)$ and $n(\hh)$ are the atomic and molecular hydrogen density, respectively; the far-UV field strength, $\iuv$; the Doppler velocity dispersion, $\delta v_D$ (or half-width at 1/e point); and the metal abundance, $Z$. $\eta$ is a beam filling factor and $\epsilon$ is a geometric 
correction factor which we describe below. 

Unlike the substructure of the fully molecular interior of clouds, which
may have many size scales, the penetration scale length of far-UV photons
and UV photochemistry insure that structures within giant molecular cloud
(GMC) complexes with UV-illuminated surfaces and CO-bearing cores have a
minimum size determined by their density and metallicity.  Measurements
and theoretical arguments show PDR structures outside the dense cores
of molecular clouds are clumpy on a size scale of $\sim 1$ pc (Burton, Hollenbach, \& Tielens 1990; Jaffe et al. 1994), comparable to or larger than the Orion beam but very much smaller than the beams in the LMC.
Since the measured intensity is a beam average over any source structure, we have to correct for the effects of different beam filling factors in order to directly compare the emitted intensities from the LMC with those from the Galactic clouds.
We define the beam filling factor, $\eta$, as the fraction of the observed beam area filled by a single cloud (not an ensemble of clouds), using the outermost edge as the cloud boundary (or the boundary between \hii\ and \hi\ regions):
\begin{equation}
  \eta = \left( \frac{ 2 R_{cloud} }{ d\ \theta_{ED} } \right)^2 \,
\end{equation}
where $R_{cloud}$ is the radius of the cloud to the outer (\hii/H) boundary, $d$ is the distance to the cloud, and $\theta_{ED}$ is the diameter of the telescope equivalent disk in radians.
$\eta$ is same for all types of emission in the same cloud.
In order to simplify model simulations in Section \ref{sec:models}, we assume that only one cloud is in the telescope beam, the telescope has a box-function beam profile, and the telescope beam area is always larger than the cloud size, i.e., $d\ \theta_{ED}$ $\geq$ $2 R_{cloud}$ or $\eta$ $\leq$ 1.

It may be more realistic to consider three dimensional 
rather than planar clouds. 
In case of a three dimensional cloud with an external far-UV field whose intensity is uniform on the surface of the cloud, the far-IR, [\cii], and \hh\ emission arise in the outer shells of the cloud, while the \twco\ \jone\ emission arises from the surface of the CO core inside the cloud.
We use a ``geometric correction parameter'', $\epsilon_{i}$ or $\epsilon_{FIR}$, in Equation \ref{eq:intensities} to simulate the {\it observed} intensity of a three dimensional cloud.
The spherical geometry parameter accounts for limb-brightening and for the size differences between the CO, [\cii], \hh, and far-IR emission regions in a given spherical cloud.

When examining our three dimensional model clouds, we can eliminate the beam filling factor, $\eta$, by dividing $I_i$ by \ifir:
\begin{equation}
  \frac{ I_{i} }{ \ifir } = 
    \frac{ \epsilon_{i} }{ \epsilon_{FIR} }\
    \frac{ f_{i}(n_H, \iuv, \delta v_D, Z) }{ 4.26 \times 10^{-4}\, \iuv }\ .
    \label{eq:Iratio}
\end{equation}
We will therefore divide the observed intensities by the far-IR intensity at each position as a means of removing distance-related beam filling factor effects in our subsequent analysis of the data from the Galaxy and the Magellanic Clouds. 

\subsection{ Relationship between $\ico$, $\icii$, and $I_{H2}$ }

Table~\ref{tbl-8} shows the average ratios between the observed \hh\ \vone, \twco\ \jone, and \cii\ intensities, and the standard deviations of the ratio distributions in the Galactic clouds and in the Magellanic Clouds.

Figures \ref{fig-08}, \ref{fig-09}, and \ref{fig-10} show plots of $\ihh/\ifir$ versus $\icii/\ifir$,
$\icii/\ifir$ versus $\ico/\ifir$, and
$\ihh/\ifir$ versus $\ico/\ifir$,
for the Galactic clouds and for the clouds in the Magellanic Clouds.
From these figures and Table~\ref{tbl-8}, the line ratios of $\log(\ihh/\icii)$ between the Galaxy and the Magellanic Clouds are in good agreement, and the ratios of $\log(\icii/\ico)$ and $\log(\ihh/\ico)$ in the Magellanic Clouds are slightly higher (but in agreement within the standard deviations) than those in the Galaxy.
Section~\ref{sec:models} will discuss our PDR models and compare the observed data with PDR models with various input parameters.

\section{ Models } \label{sec:models}

Using plane-parallel codes, the metallicity dependence of PDR structure and of emergent line intensities has been calculated and discussed by several authors (Maloney \& Black 1988; van Dishoeck \& Black 1988b; Wolfire, Hollenbach, \& Tielens 1989; Maloney \& Wolfire 1997).
The emergent intensities of the \hh\ \vone, [\cii], and \twco\ \jone\ lines from PDRs depend on $n_H$, \iuv, $\delta v_D$, and $Z$ (see Section \ref{sec:dividefir}).
As long as the \twco\ \jone\ line is optically thick, the resulting intensities of \twco, [\cii], and \hh\ \vone\ lines are not very sensitive to the metallicity. 
A simplified analysis can illustrate the reason for this insensitivity.
In the outer part of the PDR, gas-phase carbon is in the form of \cplus.
When the metallicity is lowered, the number density of \cplus\ ions drops, but the far-UV photons penetrate deeper into the clouds because the dust-to-gas ratio is also lower.
Thus, the column density of \cplus\ in the PDR is almost independent of the metallicity.
The [\cii] line is optically thin, and the intensity is proportional to the column density of \cplus.
The \twco\ \jone\ intensity depends much more weakly on the CO column density because the line is optically thick and most of emission arises on the surface of the CO region inside the cloud. 
Therefore, the \twco\ \jone\ intensity outside of very high column density cloud cores does not depend strongly on the metallicity when one considers emission from a plane-parallel slab.
We will discuss the metallicity dependence of \hh\ in Sections \ref{sec:self-shielding} and \ref{sec:without_epsilon}.

As we discussed in Section \ref{sec:dividefir}, in the case of three dimensional clouds, we use the parameter, $\epsilon_i$ or $\epsilon_{FIR}$, to account for effects such as limb-brightening and the size differences between the outer shells from which far-IR, [\cii], and \hh\ \vone\ emission arises and the inner CO core from which \twco\ \jone\ emission arises.
The depths from the cloud surface to the \cplus/C transition layer and to the H/\hh\ transition layer are about inversely proportional to the dust abundance, so the metal and dust abundance are more important parameters in spherical-shell PDRs than in plane-parallel models. 
St\"{o}rzer, Stutzki, \& Sternberg (1996) and Mochizuki (1996) modeled a
PDR on the surface of a spherical cloud with Galactic metallicity.
This Section presents our own models of spherical-shell PDRs and calculate the emergent line intensities for a range of densities, UV fields, and metallicities.

\subsection{ Codes }

We first ran a plane-parallel PDR code (van Dishoeck \& Black 1986; Black \& van Dishoeck 1987; van Dishoeck \& Black 1988a; Jansen et al. 1995) with a range of densities ($n_H$ = $5 \times 10^2$, $5 \times 10^3$, and $5 \times 10^4\ \cmv$), UV fields ($I_{UV}$ = $10$, $10^2$, $10^3$, and $10^4$), and metallicities (for the Galaxy, the LMC, and the SMC).
In this code, one side of the model cloud is exposed to UV radiation, and the cloud is divided into 200 slabs, each of which is in chemical steady state.
Since the code only includes inelastic collisions of \hh\ within v=0, the ro-vibrational level populations of \hh\ are not correct at $n_H$ $\geq$ $5 \times 10^4$ \cmv.
The PDR code calculates the gas temperature and chemical abundances in each slab. Since there is no detailed information on the grain properties in the LMC
and SMC, we assume that the formation rate of \hh\ on grains scales linearly
with the dust abundance (section \ref{sec:self-shielding}) and that the 
heating efficiencies of the grains are the same as in the Galactic case.
In practice, the lower \hh\ grain formation rate is accomplished in the models 
by decreasing the parameter $y_f$ (Black \& van Dishoeck 1987) from 1.0 to 0.25 and 0.1 in the LMC and SMC, respectively.
Table~\ref{tbl-9} lists input parameters for the code.

We employ spherically symmetric cloud models to study the effects arising in three dimensional chemical structures.
The outputs, e.g., chemical abundances and kinetic temperatures, from the plane-parallel code are applied to the spherical shells of the cloud used in the radiative transfer model.
We map the temperatures and abundances derived at each distance from the \hii/\hi\ interface by the plane-parallel model into the spherical radiation transfer model, ignoring any changes in the chemistry or thermal balance due to the difference in geometry.
The Monte Carlo code of Choi et al. (1995; hereafter MC) calculates the level populations of the atoms and molecules.
The MC code simulates photons in a one-dimensional (spherical) cloud, and adjusts the level populations according to the result of simulations until the populations converge.
We assume a purely turbulent velocity field with a Doppler velocity dispersion of 1 \kms\ ($\delta v_D$ or half-width at 1/e point) with no systematic motion.
Since the MC calculation only includes 40 slabs, the 200 slab plane parallel model was smoothed in such a way as to retain high resolution at the transition regions (Li 1997).

We use the output of MC to calculate emission line profiles using a Virtual Telescope code (Choi et al. 1995; hereafter VT).
The VT code convolves the integrated emission from each spherical shell along the line-of-sight with a virtual telescope beam profile to simulate observations. While it would be better to examine the geometric effects
using fully self-consistent models which calculate the detailed \hh\
excitation, \twco\ photodissociation, and \twco\ and \cii\
line intensities, our present three-step procedure should be good enough
to allow us to analyze the global properties of Galactic and LMC clouds
and to establish relative trends. 

\subsection{ Effect of \hh\ Self-Shielding } \label{sec:self-shielding}

Inside neutral clouds, the external far-UV field is attenuated by dust absorption, and by C and \hh\ absorption.
CO absorption of far-UV is negligible in the outer part of the cloud because most of the carbon is ionized.
\hh\ can survive in the outer parts of the cloud, either as a result of self shielding or as a result of shielding by dust.
We can analyze the conditions under which \hh\ self-shielding from far-UV photons is dominant over shielding by dust (adapted from Burton, Hollenbach, \& Tielens 1990).
Within a plane-parallel cloud, the \hh\ formation rate, $F(x)$, at depth $x$ measured from the surface of the cloud toward the center is
\begin{equation} \label{eq:formation}
  F(x) = n(\hatom, x)\, n_H\, q\, \Re\ ,
\end{equation}
where $n_H$ is the hydrogen number density, $n_H$ = $n(\hatom, x)$ + $2 n(\hh, x)$, assumed to be constant over the cloud; 
$n(\hatom, x)$ and $n(\hh, x)$ are number densities, at $x$, of H atoms and \hh\ molecules respectively;
$\Re$ is the value of the \hh\ formation rate coefficient (a quantity which depends linearly on the dust abundance) for the Galactic dust abundance;
$q$ is the dust abundance normalized to the Galactic value, i.e., $q^{GAL}$ = 1.
$\Re$ is a slowly varying function of the gas temperature ($\Re$ $\propto$ $T^{1/2}$), so we take an average value ($3 \times 10^{-17}$ cm$^{-3}$~s$^{-1}$, Burton, Hollenbach, \& Tielens 1990) for this analysis.
If we assume that, at the position we are considering, the dust optical depth is negligible and the \hh\ absorption is governed by the square root portion of the curve of growth (Jura 1974) the far-UV field is attenuated by $N(\hh,x)^{1/2}$, where 
\begin{equation} \label{eq:N_hh}
  N(\hh, x) = \int_{0}^{x} n(\hh, x)\, dx\ .
\end{equation}
The \hh\ destruction rate, $D(x)$, at depth $x$ is
\begin{equation} \label{eq:destruction}
  D(x) = \frac{\iuv\, I_{\circ}\, \beta}{N(\hh,x)^{1/2}}\, n(\hh,x)\ ,
\end{equation}
where $I_\circ$ is the unshielded dissociation rate of \hh\ at \iuv\ = 1 ($7.5 \times 10^{-11}$ s$^{-1}$, Black \& van Dishoeck 1987); and
$\beta$ is the self-shielding parameter ($4.2 \times 10^5$ cm$^{-1}$, Jura 1974).
We can integrate the steady state equation, $F(x)$ = $D(x)$, over $x$:
\begin{equation}
  n_H\, q\, \Re\, \int_{0}^{x} n(\hatom,x)\, dx 
  = \iuv\, I_{\circ}\, \beta\, \int_{0}^{x} 
    \frac{ n(\hh,x) }{ N(\hh,x)^{1/2} }\, dx\ .
\end{equation}
By substituting Equation \ref{eq:N_hh} into the above equation, we get:
\begin{equation} \label{eq:form_dest}
 n_H\, q\, \Re\, N(\hatom,x)
 = 2\, \iuv\, I_{\circ}\, \beta\, N(\hh,x)^{1/2}\ .
\end{equation}
\hh\ self-shielding is more important than dust shielding, when the far-UV field attenuation by dust is still negligible,
\begin{equation} \label{eq:tau_condition}
  \tau_{dust}(x_\circ) \leq \frac{1}{2}\ ,
\end{equation}
at the point, $x_\circ$, where the molecular hydrogen column density becomes equal to the atomic hydrogen column density,
\begin{equation} \label{eq:N_H_condition}
  N(\hatom, x_\circ) = 2 N(\hh, x_\circ)\ .
\end{equation}
$\tau_{dust}(x)$ is the optical depth of dust at $\lambda$ = 100~nm.
Even though the ratio of $\tau_{dust}/\av$ depends on the chemical composition and the size distribution of the dust, the change of this ratio from the Galaxy to the LMC and the SMC is negligible compared to that of the dust-to-gas ratio, e.g., the ratios of ${\rm A_{0.1\mu m}/A_{B}}$ in the Galaxy, the LMC, and the SMC are 3.4, 4.1, and 5.1, respectively (Pei 1992), while $\av/N_H$ changes by a factor of 10 (Table 9).
To simplify our analysis, we assume that the ratio of $\tau_{dust}/\av$ is constant and that only the dust-to-gas ratio (adopted from Bohlin et al. 1983 and Black \& van Dishoeck 1987) depends on the metallicity: 
\begin{equation}
  \tau_{dust}(x) \simeq 3.0 \av(x)\ , \label{eq:tau_dust}
\end{equation}
and
\begin{equation}
  \av(x) = 6.29 \times 10^{-22} \rho_{dust} N_H(x)\ , \label{eq:rho_dust}
\end{equation}
where $\rho_{dust}$ is the dust-to-gas ratio normalized to the Galactic value (see Table 9), and $N_H(x)$ is the hydrogen column density, $N_H(x)$ = $N(\hatom, x)$ + $2 N(\hh, x)$.
Substituting Equation \ref{eq:N_H_condition} into Equation \ref{eq:form_dest}, 
we obtain the following relationship between the column density at which self shielding becomes effective, and the density, metallicity, and strength of the incident UV field:
\begin{eqnarray} 
  N_H(x_\circ)^{1/2} 
    &=& 2\, \frac{\beta\, I_\circ}{\Re} \frac{\iuv}{q\, n_H} \nonumber \\
 &\simeq& 2.1 \times 10^{12}\, \frac{\iuv}{q\, n_H}\ . \label{eq:N_H_result}
\end{eqnarray}
From Equations \ref{eq:tau_condition}, \ref{eq:tau_dust}, \ref{eq:rho_dust}, and \ref{eq:N_H_result}, we derive the conditions under which the \hh\ self-shielding is dominant over shielding by dust:
\begin{equation}      \label{eq:self-shielding}
  \frac{ n_H }{ \iuv } \geq 1.3 \times 10^2\, \rho_{dust}^{-1/2}\ , 
\end{equation}
where we assume that the optical absorption properties and the efficiency for \hh\ formation of the dust per unit H atom vary in the same way as with dust abundance: $q = \rho_{dust}$.
Figure \ref{fig-11} shows the linear size of the \cplus\ region (\xcplus) and the \hhs\ region (\xhhs, \hhs\ denotes vibrationally excited \hh) from the PDR model results:
\xcplus\ and \xhhs\ are the distances from the surface of the cloud to the inner edges of the \cplus\ region where $n(\cplus, \xcplus)$ = $n(\twco, \xcplus)$, and of the \hhs\ region where $n(\hatom, \xhhs)$ = $n(\hh, \xhhs)$ respectively. 

The plot of \xhhs\ in Figure \ref{fig-11} shows how \hh\ self-shielding affects the depth of the \hhs\ region from the surface of the cloud.
Based on Equation \ref{eq:self-shielding}, the dotted lines divide the (\iuv, $n_H$, \xhhs) space into a region where dust absorption is dominant and a \hh\ self-shielding dominant region.
We can predict the behavior of \xhhs\ as a function of \iuv, $n_H$, and $\rho_{dust}$, using the relations derived in this section.
When dust absorption is more important than \hh\ self-shielding, the \hhs\ zone ends when the far-UV field is attenuated to a fixed level, independent of the incident field.
As a result, changes in \iuv\ result in a variation of $\tau_{dust}$ according to:
\begin{equation}
  \iuv\, \exp( -\tau_{dust} ) = C\ ,
\end{equation}
where $C$ is a constant.
Substituting Equations \ref{eq:tau_dust} and \ref{eq:rho_dust} into the above equation gives that
\begin{equation}
  N_H(\xhhs) \simeq 1.22 \times 10^{21}\, 
    \frac{ \log \iuv - \log C }{ \rho_{dust} }\ , 
\end{equation}
where $N_H(\xhhs)$ is the hydrogen column density at \xhhs.
In other words, if we increase \iuv\ by an order of magnitude, $N_H(\xhhs)$ increases only by an additional additive factor of $1.22 \times 10^{21}$  $\rho_{dust}^{-1}$ \cma.

When \hh\ self-shielding is dominant, from Equation \ref{eq:N_H_result},
\begin{equation}
  \log N_H(\xhhs) \simeq 24.6 + 2 \log \left( \frac{\iuv}{q\, n_H} \right)\ , 
\end{equation}
e.g., if \iuv\ decreases by an order of magnitude, $N_H(\xhhs)$ decreases by two orders of magnitude.
The above equation explains, why, when \hh\ self-shielding dominates, the depth of the \hhs\ shell decreases rapidly as \iuv\ decreases or as $n_H$ increases (see Figure \ref{fig-11}).

In the LMC, $(\rho_{dust}^{LMC})^{-1/2}$ $\simeq 2.2$ (see Equation \ref{eq:self-shielding}), and the dotted line is shifted toward lower \iuv\ by only a factor of 2.2.
Even in the case of the SMC, $(\rho_{dust}^{SMC})^{-1/2}$ $\simeq 3.2$.
Variations in metal abundance, therefore, do not significantly affect the \hh\ self-shielding criterion.

\subsection{ Emission Intensities without Spherical Geometry Effects, $\epsilon$ }
    \label{sec:without_epsilon}

We first ran the MC and VT codes to obtain results where the spherical natures of the model clouds, e.g., limb-brightening and the geometrical size differences (see Section \ref{sec:dividefir}), are not important by setting the virtual telescope beam size ($\theta_D$) much smaller than the cloud size ($2 R_{cloud}$).
This permits us, in effect, to obtain the line intensities from a plane-parallel cloud with a finite thickness whose front and back surfaces are exposed to external far-UV radiation.
The surfaces of the cloud are perpendicular to the line-of-sight.
These models are equivalent to setting $\eta$ = 1 and $\epsilon_{i}$ = 1 in Equation \ref{eq:intensities}: $I_i$ = $f_i$.
The resulting [\cii] intensity is from both the front and back surfaces, because the [\cii] line emission is nearly optically thin (Stacey et al. 1991).
The \twco\ \jone\ line is optically thick, so the resulting \twco\ emission comes predominantly from the front surface.

The VT code does not calculate the ro-vibrational lines of \hh.
We derive the \hh\ \vone\ emission from the \hhs\ column density, $N(\hhs)$, which results from the plane-parallel PDR model.
The electronically excited \hh\ molecules relax radiatively to the ground
electronic state and then cascade through the vibrational energy levels 
by emitting ro-vibrational lines.
In this process, the \hh\ line ratios are insensitive to $n_H$ and \iuv, and we can use a constant conversion factor between $N(\hhs)$ and \hh\ \vone\ intensity (Black \& van Dishoeck 1987):
\begin{equation} \label{eq:f_H_2}
  f_{H_2}^{front} = 2.67 \times 10^{-21}\, N(\hhs)\ \ergscmsr\ ,
\end{equation}
where $f_{H_2}^{front}$ is the intensity from the front surface of the cloud.

While the ro-vibrational lines of \hh\ are optically thin, the emission from the back surface, $f_{H_2}^{back}$, is affected by the extinction, \ak, through the cloud itself.
Using \ak\ = $0.112 \av$ (Rieke \& Lebofsky 1985) and Equation \ref{eq:rho_dust}, we estimate the observed intensity from the back surface:
\begin{equation} \label{eq:ih2back}
  \log(f_{H_2}^{back}) = \log( f_{H_2}^{front} ) 
    - \frac{ \rho_{dust} N_H }{ 3.5 \times 10^{22} \cma }\ .
\end{equation}
As we will discuss in Section \ref{sec:with_epsilon}, the cloud size (in units of hydrogen column density $N_H$) has a lower limit to keep the CO molecules from being dissociated completely in an intense far-UV field (\iuv\ $>$ $10^3$).
Figure \ref{fig-13} shows that $\rho_{dust} N_H$ should be larger than $2 \times 10^{22}\, \cma$; therefore,
\begin{equation} 
  f_{H_2}^{back} < 0.3\, f_{H_2}^{front}\ .
\end{equation}
We will neglect the \hh\ emission from the back side of the cloud in the following discussion: $f_{H_2}$ $\simeq$ $f_{H_2}^{front}$.
In some cases, there may be a bit of additional flux from the back side which can penetrate through the thinner parts of the cloud. 

Figure \ref{fig-12} shows the [\cii], \hh\ \vone, and \twco\ \jone\ emission intensities from the PDR code and the MC/VT code for the two sided planar clouds.
When dust absorption dominates (Equation \ref{eq:self-shielding}), the \hh\ intensity increases as $n_H$ increases.
On the other hand, when \hh\ self-shielding dominates, the \hh\ intensity increases as \iuv\ increases.
In the LMC model (the right plot in Figure \ref{fig-12}), the \hh\ intensity is enhanced by a factor of $10^{0.1}-10^{0.3}$ over the intensity in the Galactic model with the same $n_H$ and \iuv.
The enhancement is mainly due to the different \hh\ self-shielding criteria in the Galactic model and in the LMC model.

\subsection{ Model Emission Intensities Including the Effects of Spherical Geometry }
    \label{sec:with_epsilon}

In order to understand the effect of limb-brightening and of differences in the physical sizes of the \cplus, \hhs, and CO zones in spherical clouds of varying metallicity, we ran the VT code, setting the virtual telescope beam size ($\theta_{ED}$) to match the cloud size ($2 R_{cloud}$).
The resulting intensities are equivalent to setting $\eta$ = 1 in Equation \ref{eq:intensities}: $I_i$ = $\epsilon_i f_i$.
In Figure~\ref{fig-13}, we plot the model values of \icii, \ihh, and \ico\ versus cloud size.
In Figures \ref{fig-06}, \ref{fig-09}, and \ref{fig-10}, we overlay the results on the observed data.
\icii\ and \ico\ are obtained directly from the VT code, and the far-IR continuum emission and \ihh\ are from Equation \ref{eq:intensities} with $\epsilon_{FIR}$ and $\epsilon_{H_2}$ derived as follows: 

As we discussed in Section \ref{sec:dividefir}, the effects of spherical geometry include both limb-brightening and size differences between the different emission regions in the cloud.
The far-IR emission has very low optical depth and the \hh\ \vone\ emission is optically thin.
The far-IR and \hh\ emission regions lie on the surface of the cloud and fill the telescope beam.
We can analyze the spherical geometry effects for optically thin lines by projecting the three dimensional emission shell onto a two dimensional emission disk:
\begin{eqnarray}
  \epsilon_i &=& \frac{1}{ 2\, X_i\, \pi R_{cloud}^2 }\,
		\int_{R_{cloud}-X_i}^{R_{cloud}} 4 \pi R^2\, d R\, \nonumber \\
             &=& 2 - 2 \frac{X_i}{R_{cloud}} + \frac{2}{3} 
                 \left( \frac{X_i}{R_{cloud}} \right)^2\ , \label{eq:eps1}
\end{eqnarray}
where $X_i$ is the depth from the cloud surface to the transition regions, e.g., \xcplus\ and \xhhs\ which are defined in Section \ref{sec:self-shielding}; and $R_{cloud}$ is the radius of the cloud.
In the above equation, we assume that $n(\hhs)/n_H$ and $n(\cplus)/n_H$ are constant in the corresponding shells.
For the far-IR emission, if we consider that the incident far-UV energy is conserved as the output far-IR energy (see Equation \ref{eq:G0-I_FIR}), we can write:
\begin{equation} \label{eq:eps2}
  \epsilon_{FIR} = \frac{4\pi R_{cloud}^2}{2\, \pi R_{cloud}^2} 
                 = 2\ . 
\end{equation}
We use Equation \ref{eq:eps1} to obtain $\epsilon_{H_2}$, and Equation \ref{eq:eps2} to obtain $\epsilon_{FIR}$ = 2.

The \twco\ line arises on the surface of the CO core.
Because external UV radiation provides most of the heating, the gas temperature gradient is positive, $dT/dR$ $>$ 0 ($R = 0$ at the cloud center).
As we observe an optically thick line, the edge of the projected CO disk has a higher brightness temperature than the center, and we see limb-brightening.
On the other hand, the size of the CO core ($R_{CO}$) is smaller than the cloud size ($R_{cloud}$) which is affected by \iuv\ and $n_H$.
When $R_{cloud}$ is larger than \xcplus, and \ico\ is optically thick:
\begin{equation}  \label{eq:epsilon} \label{eq:eCO_RCO}
  \epsilon_{CO} \simeq \left( \frac{ R_{CO} }{ R_{cloud} } \right)^2
  = \left( 1 - \frac{ \xcplus }{ R_{cloud} } \right)^2\ .
\end{equation}
When we compare the spherical cloud models to plane-parallel models, the change in the CO core size ($R_{CO}$) is more significant than the limb-brightening effect.
We plot the CO intensity (by setting $\eta$ = 1) resulting from the numerical code in dotted lines in Figure \ref{fig-13}.
This method also takes into account any additional effect of opacity variations on the local CO emissivity.
As $R_{CO}$ $\rightarrow$ 0, the resulting \ico\ decreases rapidly.

\subsection{ Applying the Model to the Data }

As we discussed in Sections \ref{sec:without_epsilon} and \ref{sec:with_epsilon}, the observed far-IR, [\cii], and \hh\ \vone\ line intensities depend almost exclusively on $n_H$ and \iuv, while the observed \twco\ \jone\ line intensity depends on $n_H$, \iuv, $Z$, and $R_{cloud}$.
In this section, we apply the model results to the data and estimate $n_H$, \iuv, and $R_{cloud}$ in the regions we observed.

\subsubsection{ $\ihh$ versus $\icii$ } \label{sec:hh_cii}

In Table~\ref{tbl-8}, the standard deviations of the observed $\log(\ihh/\icii)$ are smaller than those of other ratios ($0.1-0.2$ versus $0.2-0.4$), because the values of $f_{CII}$ and $f_{H_2}$ depend similarly on $n_H$ and \iuv\ (see Figure \ref{fig-12}), and in the case of spherical clouds, the [\cii] and \hh\ intensities are not very sensitive to variations in cloud size, i.e., 2/3 $<$ $(\epsilon_{CII}, \epsilon_{H_2})$ $<$ 2 (see Equation \ref{eq:epsilon} and Figure \ref{fig-13}).

In Figure \ref{fig-08}, most of the data from NGC 2024 are within the model grids at 3.2 $< \log n_H <$ 4.2, and 1.5 $< \log \iuv <$ 3.
There are positions (at $\Delta \alpha$ = $-17\arcmin$, $-16\arcmin$, and $-15\arcmin$) in NGC 2024, however, where the \ihh\ intensities lie at least a factor of 2.5 above the model grids.
At these positions, the clouds may be small enough to be transparent for \hh\ emission from the back side of the cloud (see Equation \ref{eq:ih2back}).
Based on our experience with other positions, it is extremely unlikely that the \hh\ level populations at these positions are thermalized by effects present in high density PDR's ($n_H$ $> 5 \times 10^4$ \cmv) or by shocks (see Section \ref{sec:excitation}).
The ratios of $\log(\ihh/\icii)$ in Orion are larger than those in NGC 2024. The Orion data in Figure \ref{fig-08} lie slightly below and slightly above models for $\log n_H$ = 4.7, implying that the gas density in Orion is higher than that in NGC 2024.

The mean ratios of $\log(\ihh/\icii)$ in 30 Doradus, N159, and N160 are similar to each other, suggesting that the observed clouds in the LMC have similar densities, and that collisional de-excitation does not affect the \hh\ level populations (see Table \ref{tbl-8}). 
In Figure \ref{fig-08}, most of the data in the LMC match models in the range of 3.7 $< \log n_H <$ 4.7 and 2 $< \log \iuv <$ 3.

By comparing the observed data and the model results in Figure \ref{fig-08}, we measured the median density and far-UV field for each source.
With this derived $n_H$ and \iuv, we found the \cplus\ region depth, \xcplus, using Figure \ref{fig-11}.
The values of $n_H$, \iuv, and \xcplus\ determined in this way are listed in Table \ref{tbl-10}.

\subsubsection{  $\icii$ versus $\ico$ }

Both cloud size differences and metallicity differences affect the observed \icii/\ico\ and the \ihh/\ico\ ratios.
For the clouds in NGC 2024, the results of Jaffe et al. (1994) implied that the size of the clouds ($R_{cloud}$) in the western edge zone is smaller than in the cloud proper zone (see Section \ref{sec:galactic_data}), while the \cplus\ region depth, \xcplus, in the cloud is not much different.
As we discussed in Section \ref{sec:with_epsilon}, for a given $n_H$ and \iuv, \ico\ is proportional to $\epsilon_{CO}$.
In the LMC, the dust-to-gas ratio, $\rho_{dust}^{LMC}$, is lower than in the Galaxy by the factor of $\sim 4$, and the far-UV radiation penetrates deeper into the clouds (Figure~\ref{fig-14}):
\begin{equation} \label{eq:X_rho}
    \frac{ X_{C^+}^{LMC} }{ X_{C^+}^{GAL} } 
  \simeq \frac{ \rho_{dust}^{GAL} }{ \rho_{dust}^{LMC} } \simeq 4\ .
\end{equation}
If we assume that the cloud sizes  in the Galaxy and in the LMC are same, then $\epsilon_{CO}^{LMC}$ is less than $\epsilon_{CO}^{GAL}$.
If $\epsilon_{CO}^{LMC}$ $\simeq$ $\epsilon_{CO}^{GAL}$, the cloud size in the LMC must be larger than the cloud size in the Galaxy (see Figure~\ref{fig-14}).

The \icii/\ico\ ratio is also affected by the strength of the UV field.
Mochizuki et al. (1994) compared the $\log(\icii/\ico)$ ratios in the LMC with the ratios in the Galactic plane.
Their log-ratio in the LMC, $4.4 \pm 0.3$, agrees well with our ratio in 30 Doradus, N160, and N159 ($4.2-4.8$, see Table~\ref{tbl-8}); however, their ratio in the Galactic plane, $3.1 \pm 0.5$, is much smaller than the ratio ($3.7-4.3$, see Table~\ref{tbl-8}) in Orion and NGC 2024.
We argue that the {\it Galactic plane} at $-12\arcdeg$ $< l <$ $+26\arcdeg$ has a different physical environment from active star formation regions like Orion and NGC 2024.
Most of the $I_{60\mu m}/I_{100\mu m}$ ratios at the positions in the Galactic plane are between 0.2 and 0.3 (see Figure 2 in Nakagawa et al. 1995).
Applying these ratios to Equations \ref{eq:T_dust}, \ref{eq:G0-I_FIR}, and \ref{eq:tfir_model} (see also Equation \ref{eq:I_UV-T_dust}), we can estimate that the far-UV fields in the Galactic plane are in the range of 1.5 $< \log \iuv <$ 1.7.
The far-UV field is, therefore, much lower than toward the positions we observed in Orion and NGC 2024 (1.9 $< \log \iuv <$ 3.2).
Our model results in Figure~\ref{fig-12} show that, at $\log \iuv$ = 1.6 and $\log n_H$ = 3.7, the $\log(\icii/\ico)$ ratio is 3.2, which agrees with the observed value of the Galactic plane in Mochizuki et al. (1994).
On the other hand, the far-UV fields in 30 Doradus, N160, and N159 are very high (2.0 $< \log \iuv <$ 3.8).
Even the extended regions in the LMC in which Mochizuki et al. (1994) measured the $\log(\icii/\ico)$ ratio show higher far-UV fields (1.8 $< \log \iuv <$ 2.7, derived from the $I_{60\mu m}/I_{100\mu m}$ ratios) than those in the Galactic plane.
The more intense far-UV fields, e.g., in Orion, NGC 2024, 30 Doradus, N160, and N159, dissociate more CO molecules and have deeper \cplus\ regions, \xcplus, in Equation \ref{eq:epsilon}; therefore, the \icii/\ico\ ratio becomes larger.

Figure \ref{fig-09} plots the observed data, \icii/\ifir\ versus \ico/\ifir, and overlays the results from the planar cloud models and the spherical cloud models.
For the spherical cloud models, 
the tick marks along dashed lines mark the position in the space of models
with different column densities, $N_H(2R_{cloud})$. 
The locus of each model 
comes from the value of \icii=$\epsilon_{CII} f_{CII}$ and \ico=$\epsilon_{CO} f_{CO}$ determined for that column density.
The solid lines and dotted lines show the results of the planar cloud model (\icii=$f_{CII}$ and \ico=$f_{CO}$).
Only the data from the cloud proper zone in NGC 2024 are within the planar cloud model space, and most of the observed data deviate systematically toward lower \ico/\ifir\ or higher \icii/\ifir\ from the model results. 
The models do not include spherical geometry effects (see Section \ref{sec:without_epsilon}), and Equation \ref{eq:eps1} implies that the observed \icii\ data should not be significantly affected by going from plane-parallel to spherical geometry.
Therefore, the offset of observed points from the models on the \icii\ versus \ico\ plot in Figure \ref{fig-09} is solely a result of the effect of spherical geometry on the observed \ico\ emission ($\epsilon_{CO}$).
Given the observed value of \icii/\ifir , we measured how much the observed \ico/\ifir\ is shifted from the point expected from the models for a given density and far-UV intensity on the model grids in Figure \ref{fig-09}: 
\begin{equation}
  \Delta \log (\ico/\ifir) = \log \epsilon_{CO}\ .
\end{equation}
Applying the measured \xcplus\ and $\epsilon_{CO}$ to Equation \ref{eq:epsilon}, we derived the column densities through the clouds ($N_H(2R_{cloud})$ in Table \ref{tbl-10}) under the assumption that the clouds are spherical.
For the data from the cloud proper zone in NGC 2024, we use the uncertainty in the observations to set the lower limit, i.e., $\epsilon_{CO}$ $\geq$ $10^{-0.2}$.

Madden et al. (1997) also estimated $\epsilon_{CO}$ of the nearby irregular galaxy, IC 10, using far-IR, [\cii], and \twco\ \jone\ data.
In Table \ref{tbl-10}, we list the adopted values of \iuv\ and $\epsilon_{CO}$ at positions A and B in IC 10.
Using the dust-to-gas ratio of IC 10, $\rho_{dust}^{IC10}$ $\simeq$ 0.2, and the PDR models, we estimate $N(\cplus)$ and then $N_H(2R_{cloud})$.

\subsubsection{  $\ihh$ versus $\ico$ }

Models indicate that, like \icii, \ihh\ is not affected strongly by changing from planar to spherical geometry.
Therefore, the similar offset of the data from models in Figure \ref{fig-10} is probably due to the inappropriateness of applying the plane-parallel models for predicting \ico.
As were the data in Figure \ref{fig-09}, only the data from the cloud proper zone in NGC 2024 are within the model space, and most of the observed data deviate systematically leftward from the model results. 

The data from the SMC, which are not available in Figure \ref{fig-09}, also deviate toward lower $\log(\ico/\ifir)$ or higher $\log(\ihh/\ifir)$ from the LMC model results, if we assume that the SMC models without spherical geometry effects are not very different from the LMC models.
In the SMC, the observed \hh\ emission could be higher than the model results as a result of the unreddened \hh\ emission from the back side of the cloud.
Because [\cii] data are not available and the \hh\ data are not certain, we cannot measure $\epsilon_{CO}^{SMC}$ from the \ihh\ versus \ico\ plane.
We can still estimate a lower limit on the column density of clouds
in the SMC, assuming $\epsilon_{CO}$ $\geq 10^{-1}$.
Applying the dust temperature of $10^{1.65}$ K from IRAS 60\micron/100\micron\ ratios (see Figure \ref{fig-07}) to Equations \ref{eq:G0-I_FIR} and \ref{eq:tfir_model},
\begin{equation} \label{eq:I_UV-T_dust}
  \log \iuv = -5.65 + 5 \log T_{dust}\ ,
\end{equation}
we obtain a far-UV field (\iuv) of $10^{2.6}$ at the observed positions in the SMC.
Assuming $\log n_H$ = 3.7, we get $N_H(X_{C^+}^{SMC})$ = $4.5 \times 10^{22}$ \cma.
Therefore, the central column density ($N_H$) of the clouds in the SMC should be at least $1.3 \times 10^{23}$ \cma (see Table \ref{tbl-10}).

\subsection{ Cloud Column Densities in Galactic and Magellanic Complexes}

We estimate cloud column densities in the molecular complexes in the Galaxy and in the Magellanic Clouds.
We first derive $n_H$ and \iuv\ by comparing the observed data and the model results in the \ihh/\ifir\ versus \icii/\ifir\ plot (Figure \ref{fig-08}).
We take a median value of the data in each object, except in N159 where we consider each position separately. 
Using the values of $n_H$ and \iuv\ derived from Figure \ref{fig-08}, we get \xcplus, the depth from the surface of the cloud to the \cplus-to-C-to-CO transition layer, in Figure \ref{fig-11}.
We also get $\epsilon_{CO}$, the parameter that accounts for spherical geometry, by comparing the data and models in Figure \ref{fig-09}.
When we do not have \cplus\ or \hh\ data, we assume that $\log n_H$ = 3.7, and estimated \iuv\ from dust color temperature using $I_{60\mu m}/I_{100\mu m}$ and Equations \ref{eq:T_dust} and \ref{eq:I_UV-T_dust}.
We finally derive the central column densities of clouds using Equation \ref{eq:epsilon}, and list them in Table \ref{tbl-10} (see also Figure \ref{fig-15}).
Note that the column densities derived in this way are a measure of the number of molecules between the \hii/H boundary and the cloud center, not of the overall column denisyt through the GMC complexes.

The cloud column density measured in the western edge zone in NGC 2024 is at least a factor of 5 smaller than the column density measured in the cloud proper zone in NGC 2024.
At the four data positions in N159, the variations of $n_H$ and \iuv\ are not significant, e.g., 3.7 $< \log n_H <$ 4.1, and 2.3 $< \log \iuv <$ 2.7; however, the \ico\ intensities change by a factor of $10^{1.2}$, indicating that the column density varies more than a factor of 5 within the N159 region.
The cloud sizes represented by the hydrogen column density for a given density, $n_H$ = $5 \times 10^3$ \cmv, in the LMC ($N_H^{LMC}$ = $3-15 \times 10^{22}$ \cma ), the SMC ($> 1.3 \times 10^{23}$), and IC 10 ($6-10 \times 10^{22}$ \cma ) are bigger than those in the Galaxy ($N_H^{GAL}$ = $1-4 \times 10^{22}$ \cma ).

\subsection{ Column Density and Metallicity } \label{sec:size_Z}

Table \ref{tbl-10} and Figures \ref{fig-14} and \ref{fig-15} show that the
column density and, for constant density, cloud size depend roughly linearly on metallicity and therefore the visual extinction through clouds is constant as one goes from the Galaxy to the LMC and SMC.

McKee (1989) presented a theory of {\it photoionization-regulated star formation} (see also Bertoldi \& McKee 1996).
According to his theory, the rate of low-mass star formation is governed by ambipolar diffusion in magnetically supported clouds, and the ionization of gas is mostly by far-UV photons.
He suggested that the mean extinction of molecular clouds in dynamical equilibrium is in the range of \av\ = $4-8$ mag.
Using Equation \ref{eq:rho_dust}, the range of extinction corresponds to:
\begin{equation}
  0.64 \times 10^{22} < \rho_{dust}\, N_H < 1.3 \times 10^{22}\, \cma\ .
\end{equation}
This range of column density roughly agrees with the empirical values of Galactic GMCs using \ico\ and virial theorem: $N_H$ = $4-8 \times 10^{22}$ \cma\ (Scoville et al. 1987; Solomon et al. 1987).
In regions with higher incident UV fields, the equilibrium range should move to higher column density.
Our results from the Galaxy, the LMC, the SMC, and IC 10 are also within \av\ = $4-16$ mag, and agree with the theory to within a factor of two for the range of observed incident fields.

\section{ Conclusions }

We observed \hh\ \vone, \vtwo, and \vfive\ lines from the Magellanic Clouds, and detected emission even in regions where CO had not been seen during the ESO-SEST Key Program, implying that \hh\ is present in these regions.
With deeper CO observations, we also detected weak ($0.5-5\ \kkms$) \twco\ emission in PDRs where the previous ESO-SEST Program did not detect \twco\ \jone.
We conclude that, in all regions where we could detect \hh , CO is not completely dissociated and that the \twco\ \jone\ emission is still optically thick in the Magellanic Clouds.
Our observations may have consequences for studies using CO to trace molecular mass.

Our spherical-shell cloud model demonstrates the effects of limb-brightening and differences in the physical sizes of the \cplus, \hhs, and CO zones on the observed line strengths.
The \hh, [\cii], and \twco\ line intensities from the plane-parallel cloud model are not very sensitive to the metallicity.
When geometric effects are taken into account using spherical models, 
the \hh, and [\cii] line intensities normalized to the far-IR intensity are also not very sensitive to the metallicity; however, the \twco\ intensity is proportional to the surface area of the CO core.
As we decrease the cloud size (or total column density, $N_H$), the CO intensity normalized to the far-IR intensity decreases.

We compiled data of \hh\ \vone, \co\ \jone, [\cii], and far-IR in the star formation regions in the Magellanic Clouds and in the Galaxy, and compared them with simulated line intensities from a plane-parallel cloud.
The data in the \ihh/\ifir\ versus \icii/\ifir\ plot agree with the model results, and show no significant difference between the Magellanic Clouds and the Galaxy.
We use the \ihh/\ifir\ versus \icii/\ifir\ plot to estimate $n_H$ and \iuv\ at the positions we observed.
In the \icii/\ifir\ versus \ico/\ifir\ plot and the \ihh/\ifir\ versus \ico/\ifir\ plot, the data from the western edge zone in NGC 2024 and from the Magellanic Clouds are shifted to the lower \ico\ than the model results, which can be explained as an effect of spherical geometry.
We measured the amount of the effect, and used the result to derive the column density through the clouds.

On galaxy scales, the derived average column density and therefore size of the clouds appears to increase as the metallicity of the galaxy decreases.
Our observational result, that the mean extinction of clouds is constant and independent of their metallicity, is consistent with the idea that photoionization may play a role in regulating star formation.

\acknowledgments

We thank Luke Keller, Michael Luhman, and Thomas Benedict for contributions to the Fabry-Perot Spectrometer Project, and Jonathan Elias, Brooks Gregory, and the staff of the CTIO for their assistance in setting up our instrument.
We also thank Minho Choi for providing the MC/VT code, Wenbin Li for helping calculations, Kenji Mochizuki, David Jansen, and Marco Spaans for helpful discussions of PDR models, and Harriet Dinerstein, Bruce Draine, John Black, and the referee, Marc Kutner, for critical comments on this paper.
This work was supported by the David and Lucile Packard Foundation
and by NSF grants AST 9117373 and AST9530695.

\clearpage
\footnotesize

\begin{deluxetable}{lccccccc}
\tablecaption{ \hh\ Lines and Instrument Parameters \label{tbl-1} }
\tablewidth{14cm}
\tablehead{
  \colhead{ \hh\ Line } & \colhead{ $\lambda$\tablenotemark{a} } 
& \colhead{ Date } & \colhead{ $\theta_{ED}$\tablenotemark{b} }
& \colhead{ $F$\tablenotemark{c} } & \colhead{ $m$\tablenotemark{d} }
& \colhead{ $\Delta V_{FWHM}$\tablenotemark{e} } 
& \colhead{ $\Delta V_{BW}$\tablenotemark{f}} \\
  & \micron &  & arcsec &  &  & \multicolumn{2}{c}{ \kms }
}
\startdata
\vfive & 1.61308 & 1995 Oct & 81.0 & 23.0 & 125 & 104 & 163 \nl
\vone  & 2.12125 & 1995 Oct & 81.0 & 25.4 &  94 & 125 & 196 \nl
       &         & 1994 Dec & 88.9 & 26.4 &  94 & 120 & 189 \nl
\vtwo  & 2.24710 & 1994 Dec & 88.9 & 26.4 &  89 & 127 & 199 \nl
\enddata
\tablenotetext{a}{Wavelength in air (Black \& van Dishoeck 1987)}
\tablenotetext{b}{Beam size (equivalent disk). The beam profiles are in Figure~\ref{fig-01}.}
\tablenotetext{c}{Effective Finesse of the Fabry-Perot interferometer
  including the reflectivity, parallelism, surface quality, and incident angles}
\tablenotetext{d}{Order of interference}
\tablenotetext{e}{Full-width at half maximum of the instrument profile of an extended source.}
\tablenotetext{f}{$\Delta V_{BW} = \int I(V) dV / I_{peak}$. 
  $I(V)$ is the instrument profile of an extended source.}
\end{deluxetable}

\begin{deluxetable}{lccccc}
\tablecaption{ Frequency Switching Parameters \label{tbl-2} }
\tablewidth{12cm}
\tablehead{ 
  \colhead{ Object }     & \colhead{ \hh\ Line } 
& \multicolumn{2}{c}{ON} & \multicolumn{2}{c}{OFF} \\
  &  & \colhead{ \micron } & \colhead{ \kms }
     & \colhead{ \micron } & \colhead{ \kms }
}
\startdata
SMC & \vfive & 1.61386 & $+124$ & 1.61506 & $+347$ \nl
    & \vone  & 2.12219 & $+112$ & 2.12069 & $-100$ \nl
LMC & \vfive & 1.61464 & $+274$ & 1.61584 & $+477$ \nl
    & \vone  & 2.12321 & $+261$ & 2.12171 &  $+49$ \nl
    & \vtwo  & 2.24935 & $+249$ & 2.24785 &  $+49$ \nl
\enddata
\end{deluxetable}

\begin{deluxetable}{lccc}
\tablecaption{ Object List \label{tbl-3} }
\tablewidth{12cm}
\tablehead{ 
  \colhead{ Object\tablenotemark{a} } 
& \colhead{ $\alpha_{1950}$ } & \colhead{ $\delta_{1950}$ } 
& \colhead{ Reference }
}
\startdata 
DEM S 16   & 00 43 33.4 & $-73$ 39 05 & 1 \nl
DEM S 37   & 00 46 16.2 & $-73$ 32 47 & 1 \nl
LI-SMC 36  & 00 44 50.5 & $-73$ 22 23 & 1 \nl
30 Dor   & 05 39 11.5 & $-69$ 06 00 & 2 \nl
N159/N160& 05 40 18.2 & $-69$ 47 00 & 3 \nl
\tablenotetext{a}{ (0, 0) Position }
\tablerefs{
(1) Based on the \twco\ survey of the ESO-SEST Key Program 
    (Rubio et al. 1993);
(2) The CO peak (Poglitsch et al. 1995);
(3) Center of the \twco\ map in the N159/N160 region 
  (Booth 1993; Johansson et al. 1994). Because the reference position 
  of N160 is also N159 (0, 0) position, the center of N160 is close 
  to the (-1, 7) position in this reference system.}
\enddata
\end{deluxetable}

\begin{deluxetable}{lcccccccccl}
\scriptsize
\tablecaption{ Intensities from the Magellanic Clouds \label{tbl-4} }
\tablehead{
  \colhead{ Object } 
& \colhead{ $\Delta\alpha$\tablenotemark{a} }
& \colhead{ $\Delta\delta$\tablenotemark{a} }
& \multicolumn{2}{c}{\hh\ \tablenotemark{b}}
& \multicolumn{2}{c}{CO \tablenotemark{c}}
& \colhead{ \cii\ \tablenotemark{d} } 
& \colhead{ far-IR \tablenotemark{e} }
& \colhead{ $T_{dust}$ \tablenotemark{f} }
& \colhead{ Ref. } \\
  & & 
  & \colhead{      $I$\tablenotemark{g} } 
  & \colhead{ $\sigma$\tablenotemark{g} }
  & \colhead{ $I$ } & \colhead{ $\sigma$ }
  & \colhead{      $I$\tablenotemark{g} }
  & \colhead{      $I$\tablenotemark{g} } &  & \colhead{CO;\cii} \\
  & \multicolumn{2}{c}{ arcmin } 
  & \multicolumn{2}{c}{ $10^{-6}$ }
  & \multicolumn{2}{c}{ \kkms } 
  & \colhead{ $10^{-4}$ }
  & \colhead{ $10^{-2}$ } 
  & \colhead{ K } &
}
\startdata
DEM S 16 &$+0.11$&$-0.12$&$ 3.43$&0.66& 2.19& 0.11&\nodata&0.43&49&2; \nl
         &$+0.11$&$+0.88$&$ 1.60$&0.66& 2.39& 0.11&\nodata&0.27&42&2; \nl
DEM S 37 &$+0.38$&$-3.35$&$ 0.99$&2.20
    &\nodata&\nodata&\nodata&\nodata&\nodata& \nl
         &$-0.02$&$-0.05$&$  2.4$& 1.1&0.584&0.073&\nodata&0.65&45&1; \nl
LI-SMC 36&$  -3$&$   0$&$1.33$&1.77
      &\nodata&\nodata&\nodata&\nodata&\nodata& \nl
      &$   0$&$   0$&$  3.2$& 1.3& 6.20&0.060&\nodata&0.33&42&1; \nl
30 Dor&$   0$&$-1.5$&$  4.2$& 2.7& 2.87&\nodata&2.6&14 &45& 3;3\nl
      &$-1.5$&$   0$&$  5.3$& 2.2& 3.45&\nodata&4.2&18 &51& 3;3\nl
      &$   0$&$   0$&$ 10.8$&0.93& 10.5&\nodata&8.1&35 &75& 3;3\nl
      &$+1.5$&$   0$&$  8.6$& 2.4& 2.23&\nodata&3.7&7.8 &44& 3;3\nl
      &$   0$&$+1.5$&$  3.8$& 2.9& 3.39&\nodata&2.9&6.6 &42& 3;3\nl
      &$   3$&$  -9$&$ 0.94$&0.75&\nodata& 3\tablenotemark{h} &\nodata&1.4&43& 4; \nl
      &$   0$&$  -6$&$ 1.63$&0.78&0.746&0.052&\nodata&2.6&43& 1; \nl
      &$  -6$&$   0$&$-0.28$&0.53&\nodata& 3\tablenotemark{h} &\nodata&1.2 &56& 4;\nl
 N159 &$-0.5$&$  -5$&$  2.1$& 1.5& 25.0& 0.14&\nodata&0.31&34&5; \nl
      &$  +1$&$  -5$&$ 0.94$&0.90& 46.0& 0.14&\nodata&0.32&35&5; \nl
      &$-2.5$&$   0$&$  0.5$& 1.9& 29.4& 0.14&0.77& 1.8 &31& 5;6\nl
      &$-1.5$&$   0$&$  4.5$& 1.4& 47.3& 0.14& 2.3& 7.1 &46& 5;6\nl
      &$   0$&$   0$&$ 3.23$&0.99& 7.70&0.052& 2.5& 8.2 &42& 1;6\nl
      &$  +1$&$  +1$&$  4.8$& 1.3& 19.8&0.052& 2.9& 8.7 &56& 1;6\nl
      &$   0$&$  +3$&$ 1.80$&0.72&0.634&0.052& 1.0& 1.7 &46& 1;6\nl
 N160 &$   0$&$+4.5$&$  0.8$& 1.2&\nodata& 3\tablenotemark{h} &0.49& 1.2 &37& 5;6\nl
      &$-2.5$&$  +7$&$ 1.95$&0.80& 1.59&0.073&0.77& 2.6 &36& 1;6\nl
      &$  -1$&$  +7$&$ 6.45$&0.91& 9.89&0.073& 2.0& 15  &76& 1;6\nl
      &$   0$&$  +7$&$ 2.46$&0.71& 5.82&0.073& 1.7& 6.5 &54& 1;6\nl
      &$  -2$&$  +8$&$ 1.86$&0.64& 4.25&0.073& 1.1& 2.9 &39& 1;6\nl
      &$   0$&$  +8$&$ 2.18$&0.56& 6.76&0.073& 1.1& 1.8 &34& 1;6\nl
      &$  -3$&$  +9$&$-0.21$&0.65&\nodata& 3\tablenotemark{h} &0.68& 0.97&35& 5;6\nl
      &$  -1$&$  +9$&$ 1.98$&0.59& 2.76&0.073&0.75& 1.1 &34& 1;6\nl
      &$  -2$&$ +10$&$ 1.72$&0.51& 1.64&0.073&0.55& 1.0 &46& 1;6\nl
      &$   0$&$ +10$&$ 2.05$&0.52& 3.86&0.073&0.36& 0.75&43& 1;6\nl
      &$  -1$&$ +12$&$ -0.6$&1.3&4.49&\nodata&\nodata&0.71&41& 5;\nl
\enddata
\tablenotetext{a}{Offset from the object (0, 0) position.}
\tablenotetext{b}{Measured \hh\ \vone\ intensity with the $1 \sigma$
  statistical uncertainty.}
\tablenotetext{c}{Velocity integrated \twco\ \jone\ intensity, 
  $I = \int T_{MB}\, dv$. 1 \kkms\ = $1.57 \times 10^{-9}$ \ergscmsr. }
\tablenotetext{d}{[\cii] 158 \micron\ intensity.}
\tablenotetext{e}{IRAS Far-IR continuum intensity. 
  See Equation~\ref{eq:I_FIR}.}
\tablenotetext{f}{Dust temperature deduced from the ratio, 
  $I_{60\mu m} / I_{100\mu m}$. See Equation~\ref{eq:T_dust}. }
\tablenotetext{g}{In units of \ergscmsr .}
\tablenotetext{h}{Upper limit that lies beyond the lowest contour ($T_{MB}$ = 3 \kkms) in the published CO maps (Booth 1993; Johansson et al. 1994). See also Figures \ref{fig-03} and \ref{fig-04}. }
\tablerefs{
  (1) This work; (2)Rubio et al. 1993; (3) Poglitsch et al. 1995; 
  (4) Booth et al. 1993; (5) Johansson et al 1994; (6) Israel et al. 1996.
}
\footnotesize
\end{deluxetable}

\begin{deluxetable}{lcccccc}
\tablecaption{ \hh\ Line Ratios \label{tbl-5} }
\tablehead{
  \colhead{ Object } 
& \colhead{ $\Delta\alpha$ } & \colhead{ $\Delta\delta$ }
& \multicolumn{2}{c}{\vtwo} & \multicolumn{2}{c}{\vfive} \\
  &  &  
& \colhead{ Intensity\tablenotemark{a} } 
& \colhead{ Ratio\tablenotemark{b} } 
& \colhead{ Intensity\tablenotemark{a} } 
& \colhead{ Ratio\tablenotemark{b} } \\
  & \multicolumn{2}{c}{ arcmin }
  & \colhead{ $10^{-6}$ } &  & \colhead{ $10^{-6}$ } &
}
\startdata
DEM S 16 
 &+0.11&$-0.12$&  \nodata  &  \nodata   &$0.76\pm0.34$&$0.22\pm0.11$ \nl
 &+0.11&$+0.88$&  \nodata  &  \nodata   &$0.27\pm0.36$&$0.17\pm0.24$ \nl
30 Dor
 &   0  &    0 &$4.0\pm1.5$&$0.37\pm0.14$&$1.48\pm0.37$&$0.138\pm0.036$\nl
 &$+1.5$&  0   &  \nodata  &  \nodata   &$1.42\pm0.51$&$0.166\pm0.075$\nl
N159  
 &$+1$  &$+1$  &  \nodata  &  \nodata   &$0.65\pm0.41$&$0.135\pm0.091$\nl
N160
 & $-1$ & $+7$ &$3.6\pm1.9$&$0.56\pm0.30$&$2.36\pm0.41$&$0.366\pm0.082$\nl
\enddata 
\tablenotetext{a}{Measured intensity with the $1\sigma$ statistical
  uncertainty in units of \ergscmsr.}
\tablenotetext{b}{Intensity ratio with respect to \vone\ line intensity.}
\end{deluxetable}

\begin{deluxetable}{cccccc}
\tablecaption{ Comparing with Other \hh\ \vone\ Data in N159 \label{tbl-6} }
\tablewidth{14cm}
\tablehead{
  \colhead{ Ref. } & \colhead{ Position\tablenotemark{a} } 
& \colhead{ Beam size\tablenotemark{b} } 
& \colhead{ $\Omega$\tablenotemark{c} } 
& \colhead{ Flux\tablenotemark{d} } 
& \colhead{ Intensity\tablenotemark{e} } \\
  & \colhead{ arcmin } & \colhead{ arcsec } & \colhead{ $10^{-9}$ sr }
  & \colhead{ $10^{-14}$ }
  & \colhead{ $10^{-6}$ }
} 
\startdata
1 & $(+1.11, +0.85)$ & $6 \times 6$   & 0.85 & 1.7 & 20  \nl
2 & $(+1.11, +0.85)$ & $\phi 13$      & 3.1  & 5.2 & 17  \nl
3 & $(+1.11, +0.85)$ & $10 \times 21$ & 5.0  & 8.5 & 17  \nl
4 & $(+1.0, +1.0)$   & $\phi 81$      & 121  & 58  & 4.8 \nl
\enddata 
\tablenotetext{a}{ Offset from the N159 (0, 0) position (see Table 3).
  $(+1.11, +0.85)$ corresponds to 
  ``N159 Blob'' (Heydari-Malayeri \& Testor 1985). }
\tablenotetext{b}{ $\phi 13$ denotes $\theta_{ED}$ = 13\arcsec. }
\tablenotetext{c}{ Solid angle of the beam. }
\tablenotetext{d}{ In units of \ergscm . }
\tablenotetext{e}{ In units of \ergscmsr . }
\tablerefs{
  (1) Israel \& Koornneef 1992; (2) Krabbe et al. 1991; 
  (3) Kawara, Nishida, \& Taniguchi 1988 (4) This work }
\end{deluxetable}

\begin{deluxetable}{lccccccccl}
\tablewidth{0pt}
\tablecaption{ Intensities from the Galactic Clouds \label{tbl-7} }
\tablehead{
  \colhead{ Object\tablenotemark{a} }
& \colhead{ $\Delta\alpha$ } & \colhead{ $\Delta\delta$ }
& \multicolumn{2}{c}{\hh} & \colhead{ CO } & \colhead{ \cii } 
& \colhead{ far-IR } & \colhead{ $T_{dust}$ }
& \colhead{ Ref.\tablenotemark{b} } \\
  &  &
  & \colhead{      $I$\tablenotemark{c} }
  & \colhead{ $\sigma$\tablenotemark{c} } & \colhead{ $I$ } 
  & \colhead{      $I$\tablenotemark{c} }
  & \colhead{      $I$\tablenotemark{c} } &  & \colhead{ CO;CII } \\
  & \multicolumn{2}{c}{ arcmin } 
  & \multicolumn{2}{c}{ $10^{-6}$ }
  & \colhead{ \kkms } 
  & \colhead{ $10^{-4}$ }
  & \colhead{ $10^{-2}$ } 
  & \colhead{ K } &
}
\startdata
Orion   &$  +8 $&   0 &23 & 6 &34.0&9.9 &56  &55& 1;2\\
        &$ +10 $&   0 &28 & 5 &29.0&6.1 &20  &39& 1;2\\
        &$ +12 $&   0 &5.2&1.6&29.5&4.3 &8.3 &36& 1;2\\
        &$+1.84$&+9.83&27 & 4 &60.7&4.5 &41  &42& 1;3\\
NGC 2024:Edge&$ -17$&0 &9 &1.9&13.4&5.6 &3.3 &36& 4;4\\
        &$ -16 $&   0 &12 & 3 &18.6&6.7 &4.7 &37& 4;4\\
        &$ -15 $&   0 &13 & 2 &14.4&8.6 &4.9 &37& 4;4\\
        &$ -14 $&   0 &3.3&1.2&6.70&5.0 &4.3 &37& 4;4\\
        &$ -13 $&   0 &4.5&1.8&15.6&4.7 &4.4 &35& 4;4\\
        &$ -12 $&   0 &6.4&1.9&51.0&5.6 &6.6 &38& 4;4\\
        &$ -11 $&   0 & 8 & 2 &84.6&6.3 &10  &41& 4;4\\
NGC 2024:Prop&$  -9 $& 0&9 &3 &110 &8.8 &19  &49& 4;4\\
        &$  -8 $&   0 & 6 & 3 &132 &11  &15  &43& 4;4\\
        &$  -6 $&   0 & 8 & 3 &154 &10  &26  &47& 4;4\\
        &$  -5 $&   0 &11 & 4 &130 &9.9 &55  &52& 4;4\\
\enddata
\tablenotetext{a}{
  Orion (0,0):
    {\rm $\alpha_{1950} = 5^h 32^m 49\fs0$}, 
    {\rm $\delta_{1950} = -05\arcdeg 25\arcmin 16\arcsec$} ($\theta^1$~Ori C).
  NGC 2024 (0,0):
    {\rm $\alpha_{1950} = 5^h 39^m 14\fs0$}, 
    {\rm $\delta_{1950} = -01\arcdeg 57\arcmin 00\arcsec$}.
  {\it NGC 2024:Edge} and {\it NGC 2024:Prop} have the same (0, 0) position.
  See the text for the differences between NGC 2024:Edge and NGC 2024:Prop. }
\tablenotetext{b}{ \hh\ data is from Luhman \& Jaffe 1996 }
\tablenotetext{c}{ In units of \ergscmsr . }
\tablerefs{
  (1) Schloerb \& Loren 1982; (2) Stacey et al. 1993; (3) Luhman et al. 1997;
  (4) Jaffe et al. 1994 }
\tablecomments{ See the table footnotes in Table 4 for other columns. }
\end{deluxetable}

\begin{deluxetable}{lccccccc}
\tablewidth{14cm}
\tablecaption{ Intensity Ratios between \hh\ \vone, [\cii], and \twco\ \jone
   \label{tbl-8} }
\tablehead{
  \colhead{ Object }  & \colhead{ N\tablenotemark{a} }
& \multicolumn{2}{c}{ $\log(\ihh/\icii)$ } 
& \multicolumn{2}{c}{ $\log(\icii/\ico)$ } 
& \multicolumn{2}{c}{ $\log(\ihh/\ico)$ } \\
 & & \colhead{ Avg.\tablenotemark{b} }
  & \colhead{ S. D.\tablenotemark{c} }
  & \colhead{ Avg. } & \colhead{ S. D. } 
  & \colhead{ Avg. } & \colhead{ S. D. }
}
\startdata
The Galaxy      &15 & -1.85 &  0.28 &  4.04 &  0.37 & 2.19 & 0.47 \\
...Orion        & 4 & -1.53 &  0.31 &  4.01 &  0.25 & 2.48 & 0.32 \\
...NGC 2024:Prop& 4 & -2.08 &  0.13 &  3.68 &  0.03 & 1.61 & 0.14 \\
...NGC 2024:Edge& 7 & -1.91 &  0.15 &  4.26 &  0.37 & 2.35 & 0.38 \\
The LMC         &17 & -1.70 &  0.17 &  4.37 &  0.44 & 2.71 & 0.40 \\
...30 Dor       & 5 & -1.81 &  0.11 &  4.82 &  0.14 & 3.01 & 0.23 \\
...N160         & 8 & -1.59 &  0.18 &  4.18 &  0.21 & 2.59 & 0.20 \\
...N159         & 4 & -1.78 &  0.07 &  4.19 &  0.65 & 2.41 & 0.62 \\
The SMC         & 4 &\nodata&\nodata&\nodata&\nodata& 2.89 & 0.40 \\
\enddata
\tablenotetext{a}{ Number of observed positions. }
\tablenotetext{b}{
  Average of $\log(\icii/\ihh)$.
  \ihh, \icii, and \ico\ denote line intensities of \hh\ \vone, 
  [\cii] 158 \micron, and \twco\ \jone. }
\tablenotetext{c}{ standard deviation (a measure of how widely values are dispersed from the average value) of $\log(\icii/\ihh)$. }
\end{deluxetable}

\begin{deluxetable}{lccccc}
\tablecaption{ Input parameters for the PDR Code \label{tbl-9} }
\tablewidth{14cm}
\tablehead{ 
  \colhead{ Object }
& \colhead{ $N_H$\tablenotemark{a} }
& \colhead{ $Z_C$\tablenotemark{b} }
& \colhead{ $Z_O$\tablenotemark{b} }
& \colhead{ $\rho_{dust}$\tablenotemark{c} }
& \colhead{ $\zeta$\tablenotemark{d} } \\
  & \colhead{ $10^{22} \cma$ } &  &  &  
  & \colhead{ $\rm 10^{-17} s^{-1}$ }
}
\startdata
The Galaxy &  1 & 1    & 1    & 1    & 5 \\
The LMC    &  4 & 0.25 & 0.25 & 0.25 & 5 \\
The SMC\tablenotemark{e}
           & 10 & 0.1  & 0.33 & 0.1  & 5 \\
\enddata
\tablenotetext{a}{ H column density of the model cloud. 
  $N_H$ = $N(\hatom)$ + $2N(\hh)$. }
\tablenotetext{b}{ Carbon and Oxygen abundances normalized to the 
  Galactic values: [C]$^{GAL}$ = $1.6 \times 10^{-4}$, and 
  [O]$^{GAL}$ = $5 \times 10^{-4}$.
  Because the abundances are uncertain, 
  we use simplified values in the model calculation.} 
\tablenotetext{c}{ Dust-to-gas ratio normalized to the Galactic value.
  [\av/$N_H$]$^{GAL}$ = $6.29 \times 10^{-22}$ mag/\cma\
  (Bohlin et al. 1983; Black \& van Dishoeck 1987).}
\tablenotetext{d}{ An atomic hydrogen cosmic-ray ionizing frequency.}
\tablenotetext{e}{ Only for a standard model with 
  $n_H$ = $5 \times 10^3$ \cmv and $I_{UV}$ = $10^3$.}
\end{deluxetable}

\begin{deluxetable}{lccccccc}
\tablecaption{ Derived Cloud Column Densities \label{tbl-10} }
\tablewidth{14cm}
\tablehead{
  \colhead{ Object } &\colhead{ $\rho_{dust}$\tablenotemark{a} } 
& \colhead{ N\tablenotemark{b} } 
& \colhead{ $\log n_H$ } & \colhead{ $\log \iuv$ }
& \colhead{ $N(\xcplus)$\tablenotemark{c} } 
& \colhead{ $\epsilon_{CO}$\tablenotemark{d} } 
& \colhead{ $N_H(2R_{cloud})$\tablenotemark{e} } \\
  & & & & & & & \colhead{ $10^{22} \cma$ }
}
\startdata
The Galaxy       &  1 & $>15$ &    &    &      &         &        \\
...Galactic Plane\tablenotemark{f}
                & &\nodata& (3.7) & 1.6 & 0.29 & $>0.63$ & $>1.4$ \\
...Orion         &    & 4 &   4.7 & 2.5 & 0.27 & 0.44    & 1.6    \\
...NGC 2024:Prop &    & 4 &   3.7 & 2.5 & 0.45 & $>0.63$ & $>4.3$ \\
...NGC 2024:Edge &    & 7 &   3.7 & 1.8 & 0.32 & 0.083   & 0.89   \\
The LMC          &0.25& 17&       &     &      &         &        \\
...30 Dor        &    & 5 &   4.0 & 2.7 & 1.5  & 0.048   & 3.9    \\
...N160          &    & 8 &   4.2 & 2.7 & 1.5  & 0.17    & 5.0    \\
...N159$(-1.5,0)$&    & 1 &   4.1 & 2.7 & 1.5  & $>0.63$ & $>15$  \\
...N159$(+1,+1)$ &    & 1 &   3.9 & 2.7 & 1.7  & 0.36    & 8.4    \\
...N159$(0,0)$   &    & 1 &   3.7 & 2.7 & 1.8  & 0.15    & 5.8    \\
...N159$(0,+3)$  &    & 1 &   4.1 & 2.3 & 1.2  & 0.063   & 3.2    \\
The SMC\tablenotemark{f}
                 & 0.1& 4 & (3.7) & 2.6 & 4.5  & $>0.1$  & $>13$  \\
IC 10\tablenotemark{f,g} & 0.2& 2 &    &   &      &         &        \\
...Position A    &    & 1 & (3.7) & 2.7 & 2.2  & 0.3     & 9.9    \\
...Position B    &    & 1 & (3.7) & 2.6 & 2.0  & 0.1     & 5.8    \\ 
\enddata
\tablenotetext{a}{ Dust-to-gas ratio normalized to the Galactic value. }
\tablenotetext{b}{ Number of observed positions. }
\tablenotetext{c}{ Depth from the surface of the cloud to the 
  \cplus-to-C-CO transition layers. }
\tablenotetext{d}{ Parameter of the spherical geometry effect. }
\tablenotetext{e}{ The central column density of the spherical cloud.
  Note that $N_H(2R_{cloud})$ = $2 R_{cloud} n_H$, and 
  $R_{cloud}$ = \xcplus\ + $R_{CO}$. }
\tablenotetext{f}{ We assume that $\log n_H$ = 3.7. 
  We derive $\log \iuv$ from IRAS far-IR data. 
  See the text for details. }
\tablenotetext{g}{ We adopt results from Madden et al. (1997). }
\end{deluxetable}

\clearpage
\footnotesize

\begin{figure}[t]
  \vbox to 8cm { \plotfiddle{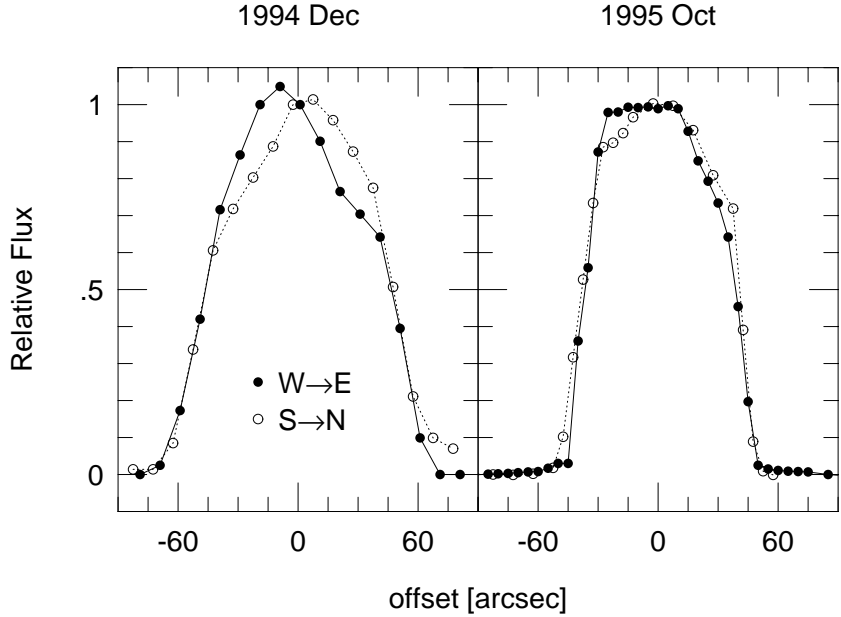} {8cm}{0}{45}{45}{-200}{-15} }
  \caption{ \label{fig-01}
    Beam profiles of the University of Texas Near-Infrared Fabry-Perot Spectrometer at the CTIO 1.5~m telescope in 1994 December and 1995 October.
}
\end{figure}

\begin{figure}[b]
  \vbox to 8cm { \plotfiddle{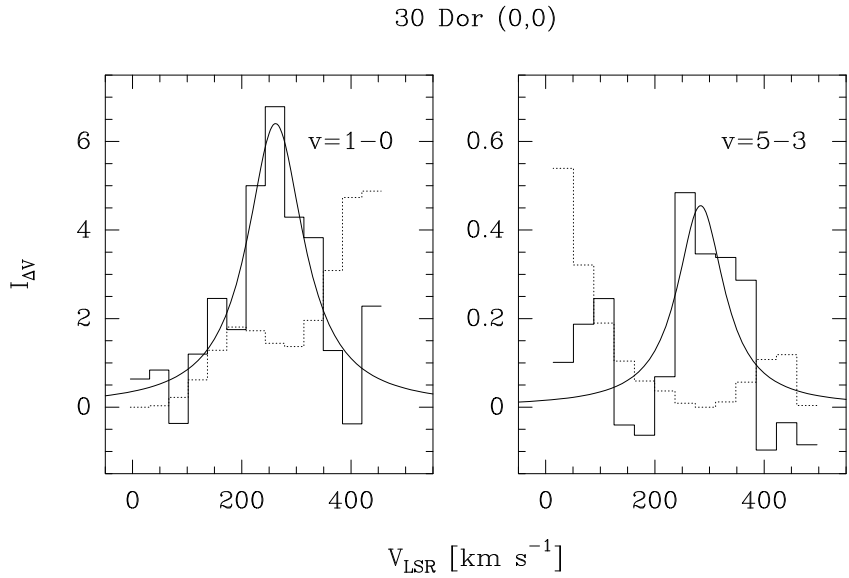} {8cm}{-90}{50}{50}{-200}{285} }
    \caption{ \label{fig-02}
    Spectra of \hh\ \vone\ and \vfive\ emission lines (solid histogram) at the 
$(0,0)$ position of 30 Doradus in units of $10^{-8}$ \ergscmsrkms. 
The smooth curve shows a fit to the spectra with the instrument parameters using {\it GaussFit} (Jefferys 1990).
The overlapped telluric OH lines (dotted line), measured at an off-source position, are scaled down by 20 on the \hh\ \vone\ spectrum and by 100 on the \hh\ \vfive\ spectrum.
  }
\end{figure}

\begin{figure}[p]
  \vbox to 11.5cm{ \plotfiddle{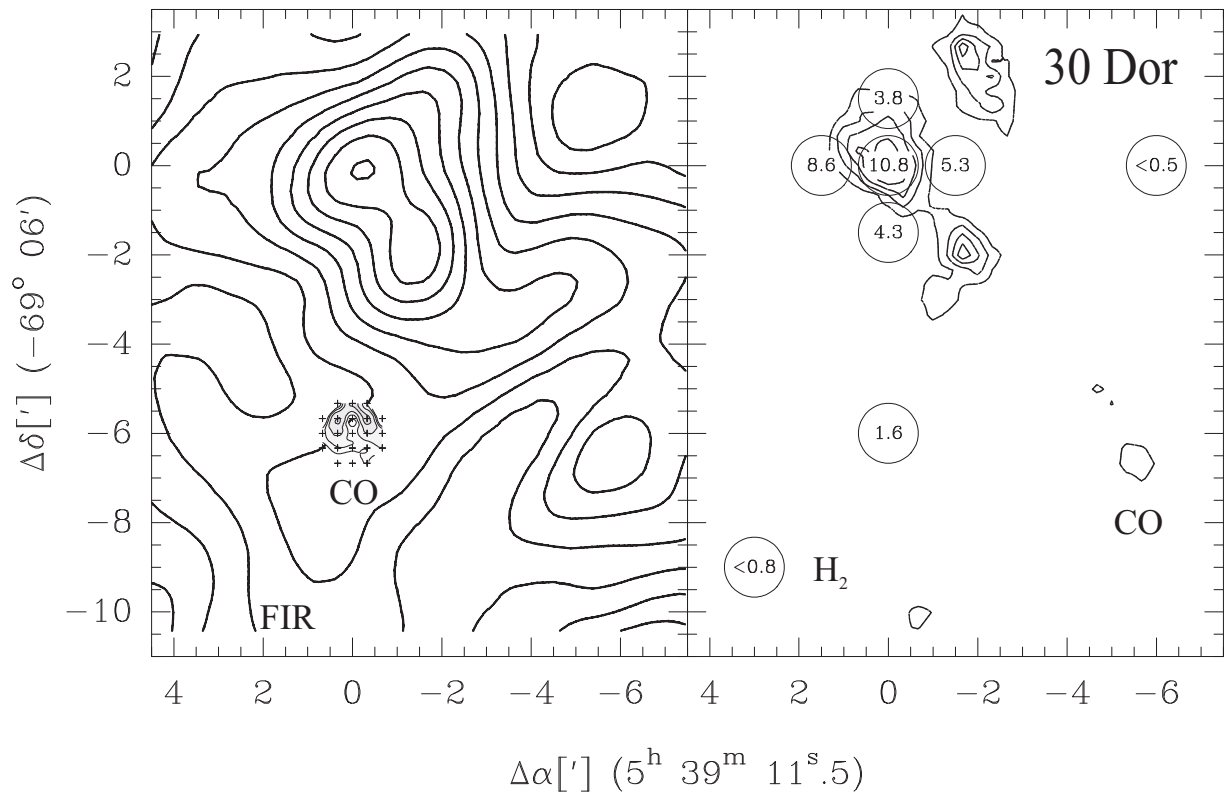}{11.5cm}{-90}{75}{75}{-300}{400} }
  \caption{ \label{fig-03}
    Intensity maps  of the 30 Doradus region.
The (0, 0) position is 
$\alpha_{1950}$ = $5^{\rm h}$ $39^{\rm m}$ $11\fs 5$ and
$\delta_{1950}$ = $-69\arcdeg$ $6\arcmin$ $0\arcsec$.
The axes show the R.A. and Dec. offset in arcmin.
    {\bf Left:} IRAS far-IR continuum map (thick lines) and new \twco\ \jone\ map (thin lines near (0, -6\arcmin)).
The far-IR intensities are calculated from the observed  $I_{60\mu m}$ and $I_{100\mu m}$ using Equation \ref{eq:I_FIR}.
The CO intensity data are obtained from this work, and the crosses show the observed positions.
The contours are spaced at logarithmic intervals:
\ifir\ = 
0.398, 0.631, 1.00, 1.58, 2.51, 3.98, 6.31, 10.0, 15.8, 25.1, 39.8 
(i.e., $10^{0.2 n}$, $n$ =$-2$, $-1$, ..., 8) 
in units of $10^{-2}$ \ergscmsr ; and
$T_{MB}$ =
0.398, 0.631, 1.00, 1.58 
(i.e., $10^{0.2 n}$, $n$ = $-2$, $-1$, 0, 1) 
in units of \kkms .
    {\bf Right:} Plotted \hh\ \vone\ data (within the circles) overlaid onto the published \twco\ \jone\ map. The numbers are in units of $10^{-6}$ \ergscmsr, and the size of the circles are the equivalent-disk size of the UT FPS beam. 
For the CO data (Booth 1993, Johansson 1994) the contours are spaced 
linearly:
$T_{MB}$ = 3, 6, 9, ... in units of \kkms .
  }
\end{figure}

\begin{figure}[p]
  \vbox to 14cm { \plotfiddle{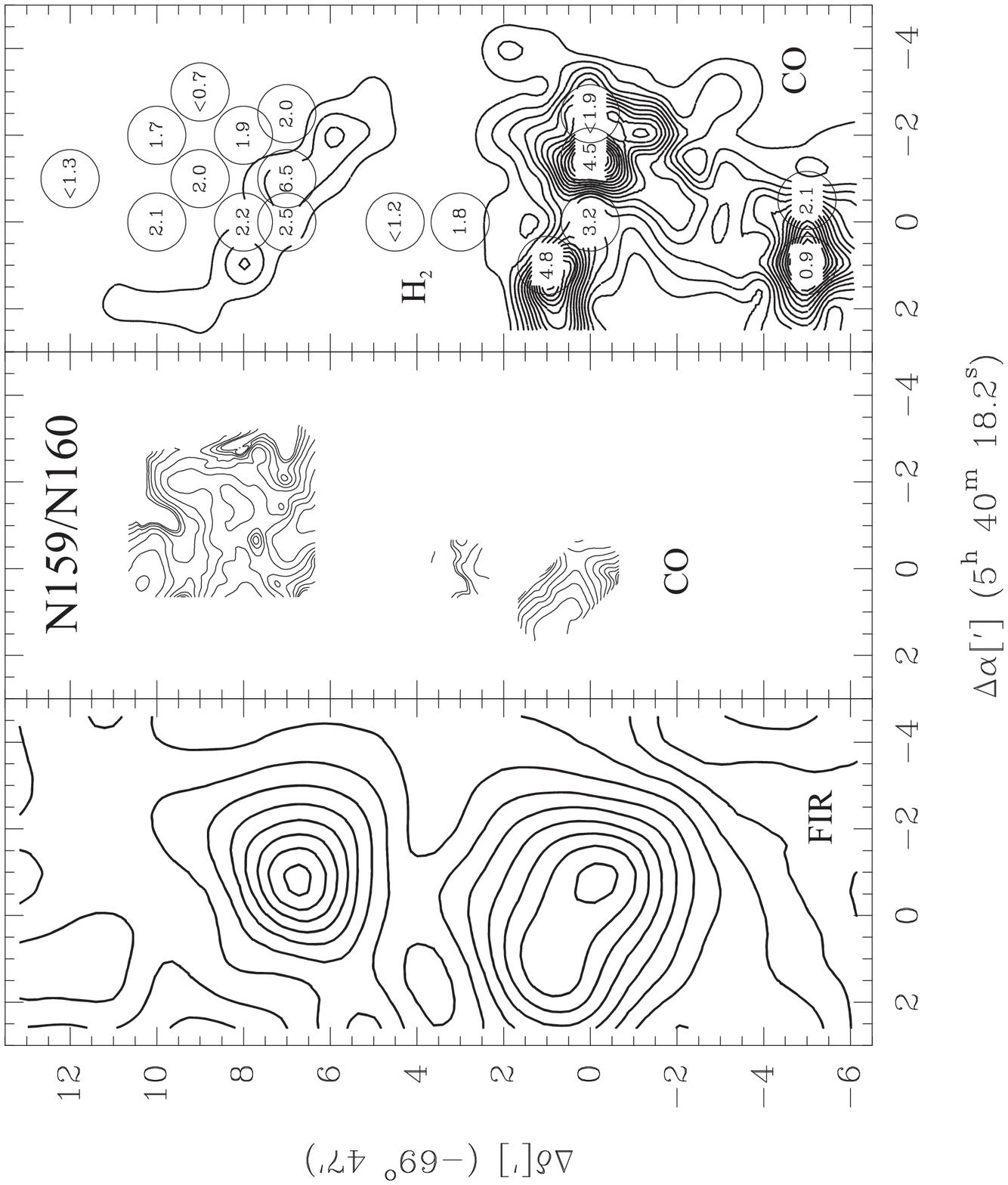} {14cm}{-90}{75}{75}{-300}{450} }
  \caption{ \label{fig-04}
    Intensity maps of the N159/N160 region.
    {\bf Left:}
Far-IR continuum map.
    The contour intervals are spaced at logarithmic intervals:
\ifir\ = 
0.158, 0.251, 0.398, 0.631, 1.00, 1.58, 2.51, 3.98, 6.31, 10.0, 15.8 
in units of $10^{-2}$ \ergscmsr. 
The far-IR peak at ($-1\arcmin, +7\arcmin$) is near N160 and the peak at ($-1\arcmin, 0$) is near N159.
    {\bf Middle:} New \twco\ \jone\ map. 
See the caption in Figure \ref{fig-05} which shows the same data.
    {\bf Right:} Plotted \hh\ \vone\ data (within the circles) overlaid onto the published \twco\ \jone\ map. See the caption for the right map of Figure \ref{fig-03}.
  }
\end{figure}

\begin{figure}[p]
  \vbox to 17cm { \plotfiddle{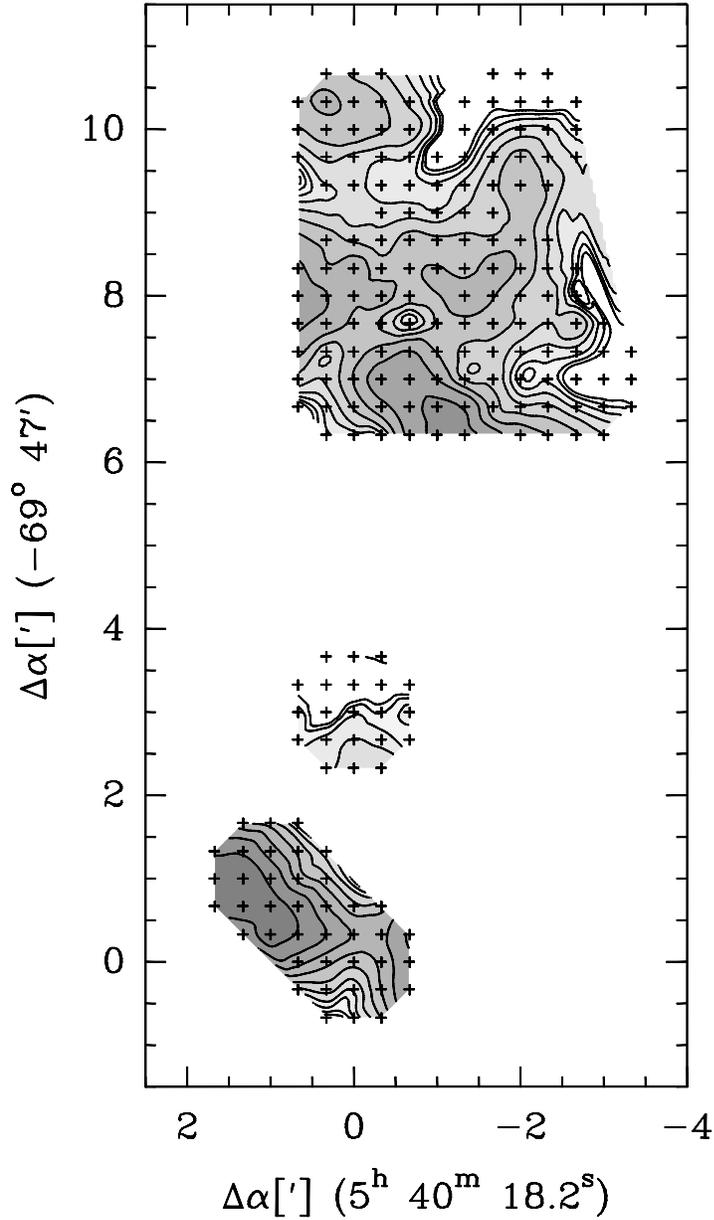} {17cm}{0}{72}{72}{-220}{-15} }
  \caption{ \label{fig-05}
    New \twco\ \jone\ intensity map of the N159/N160 region.
The contours are spaced at logarithmic intervals: 
(for the ($-1\arcmin, +8\farcm5$) region) 
$T_{MB}$ = 0.398, 0.631, 1.00, 1.58, 2.51, 3.98, 6.31, 10.0, 15.8 \kkms;
(for the ($0, +3\arcmin$) region) 
0.398, 0.631, 1.00, 1.58 \kkms;
(for the ($+0\farcm5, +0\farcm5$) region) 
0.398, 0.631, 1.00, 1.58, 2.51, 3.98, 6.31, 10.0, 15.8, 25.1 \kkms.
The observed positions are plotted in plus signs.
    }
\end{figure}

\begin{figure}[p]
  \vbox to 12cm { \plotfiddle{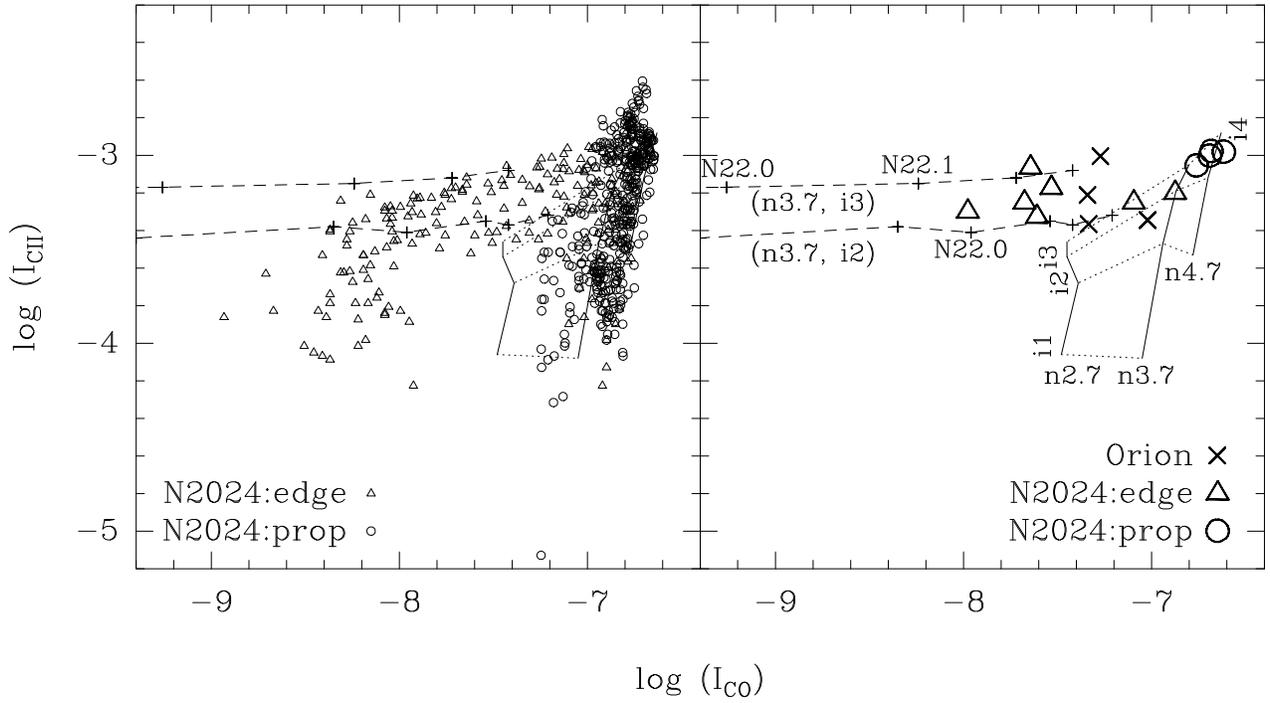} {12cm}{-90}{75}{75}{-285}{395} }
  \caption{ \label{fig-06}
    [\cii] 158 \micron\ intensity versus \twco\ J=1$\rightarrow$0 intensity 
     in units of \ergscmsr. 
    The left plot is replicated from Figure 4 in Jaffe et al. (1994). 
    The right plot shows the distribution of the Galactic data which 
    we are using in this work.
    We also overlay the results from the plane-parallel models, using the Galactic parameters. See Figures \ref{fig-08} and \ref{fig-09} for explanations.
 }
\end{figure}

\begin{figure}[p]
  \vbox to 13cm { \plotfiddle{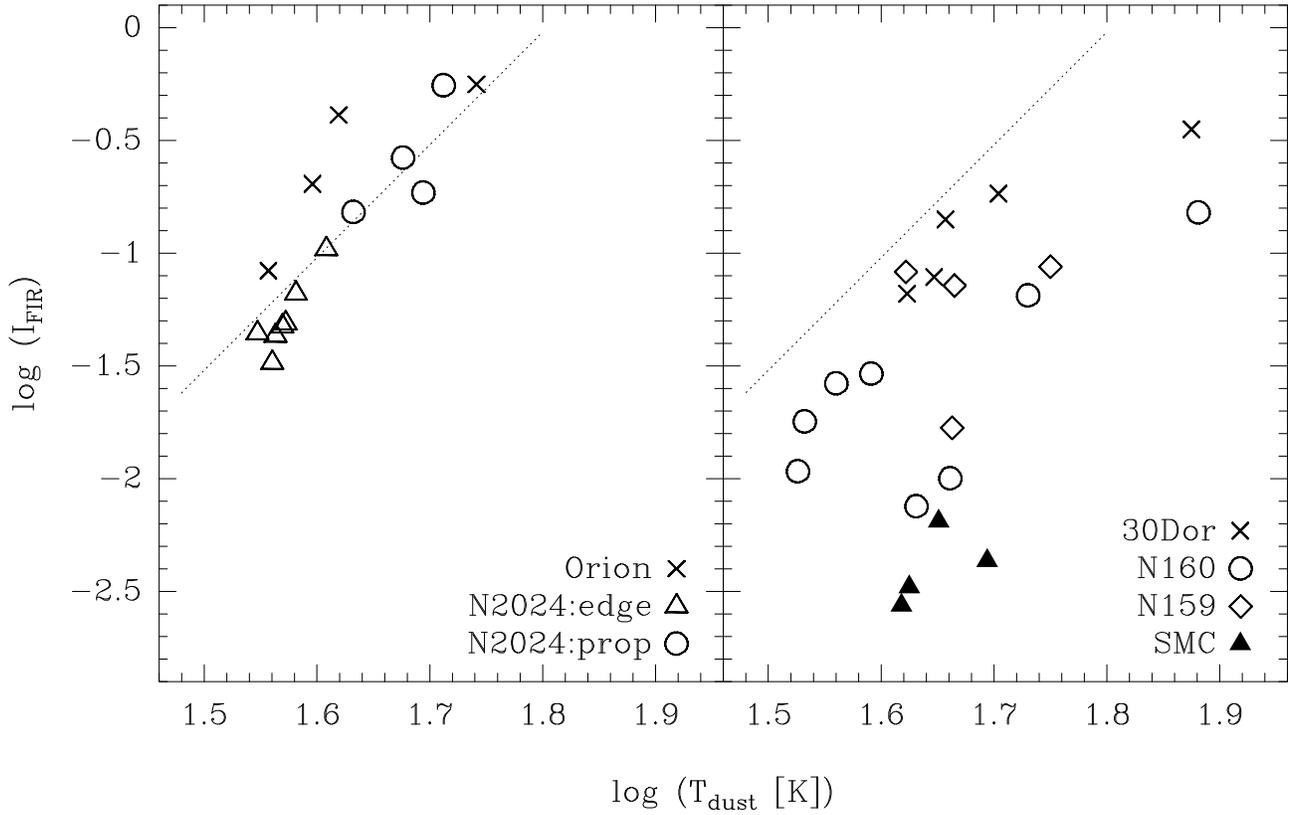} {13cm} {-90}{75}{75}{-285}{415} }
  \caption{ \label{fig-07}
Far-IR intensities (Equation \ref{eq:I_FIR} and Table-94 an6 7)  versus $T_{dust}$ (Equation \ref{eq:T_dust} and Tables 4 and 7).
    \ifir\ is in units of \ergscmsr.
    The data in the left plot are from the Galactic clouds and, in the right plot, from the LMC.
    The dotted line shows the model in Equation \ref{eq:tfir_model}. 
    We do not include the effects of spherical geometry ($\epsilon_{FIR}$, see Section \ref{sec:with_epsilon}) in the model calculation.
  }
\end{figure}

\begin{figure}[p]
  \vbox to 12cm { \plotfiddle{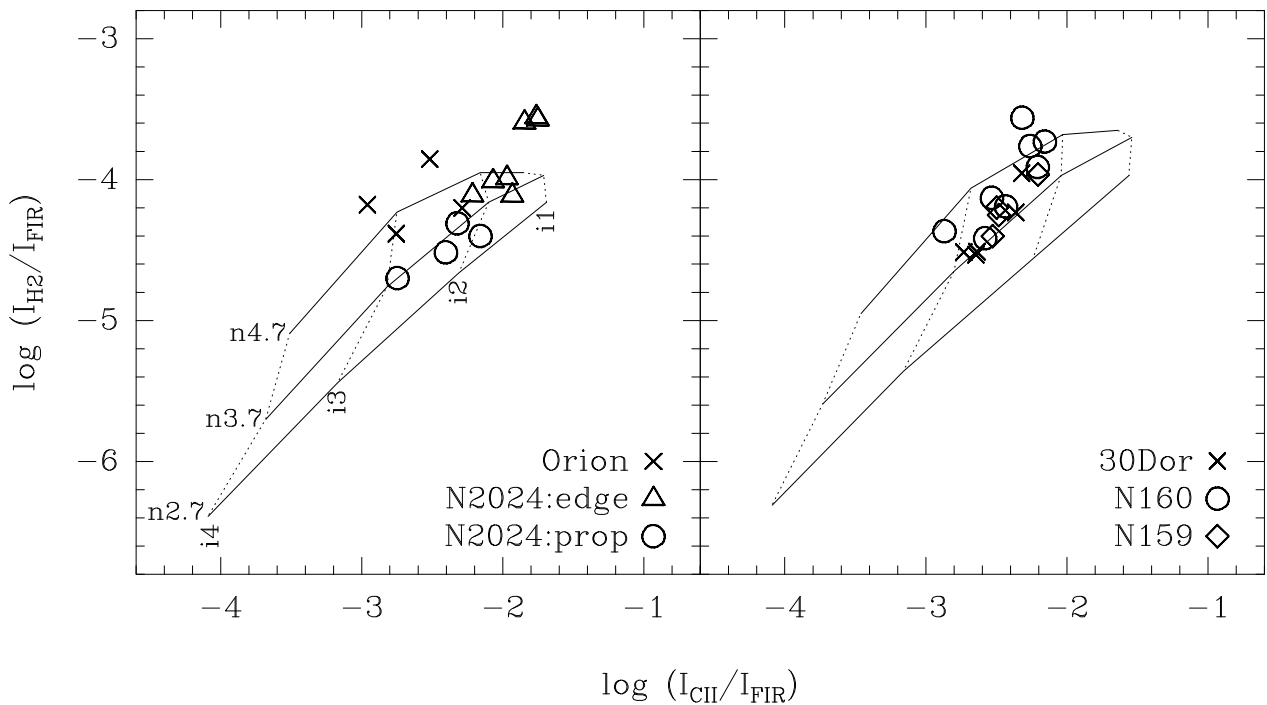} {12cm}{-90}{75}{75}{-285}{395} }
  \caption{ \label{fig-08}
     $\ihh/\ifir$ versus $\icii/\ifir$.
    The data in the left plot is from the Galactic clouds and the right plot from the LMC.
    We overlay the results from the two-sided plane-parallel models (see Section~\ref{sec:without_epsilon}).
    We used the Galactic parameters for the model results in the left plot and the LMC parameters in the right plot (see Table 9 for the list of parameters).
    In the model grids, the solid-lines and the dotted lines connect the same $n_H$ and the same \iuv\ respectively.
$n3.7$ denotes $n_H$ = $5 \times 10^3 \cmv$, and $i3$ denotes \iuv\ = $10^3$.
}
\end{figure}

\begin{figure}[p]
  \vbox to 12cm { \plotfiddle{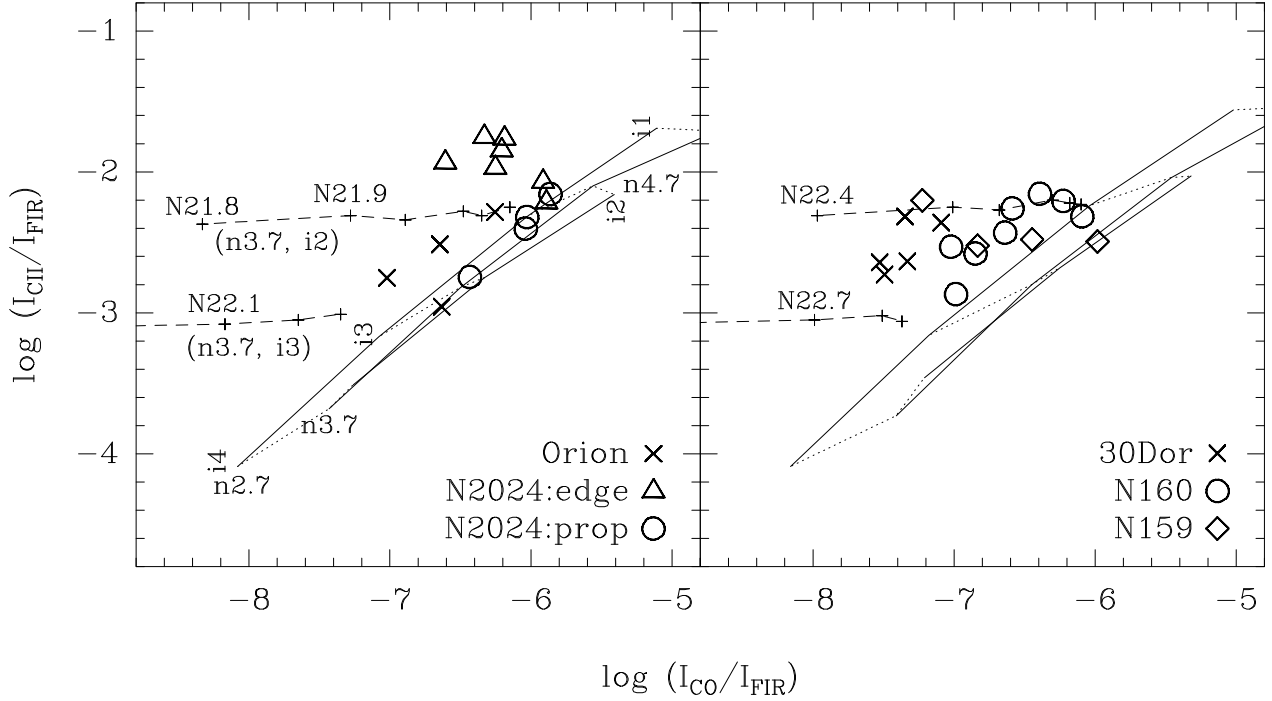} {12cm}{-90}{75}{75}{-285}{395} }
  \caption{ \label{fig-09}
    $\icii/\ifir$ versus $\ico/\ifir$.
See Figure \ref{fig-08} for the explanations of the model grids of the solid lines and the dotted lines.
We also plot models which include the effects of spherical geometry (shown as dashed lines, see Section \ref{sec:with_epsilon}).
The dashed lines connect the spherical models with the same $n_H$ and \iuv\ with different cloud sizes: 
the upper dashed line is for $n_H$ = $5 \times 10^3$ \cmv\ and \iuv\ = $10^2$; and the lower line for $n_H$ = $5 \times 10^3$ \cmv\ and \iuv\ = $10^3$.
The plus sign on the dashed line marks the cloud size spaced by 0.1 in logarithmic-scale.
$N22.3$ denotes $N_H(2R_{cloud})$ = $2 n_H R_{cloud}$ = $10^{22.3}$ \cma.
Also see Figure \ref{fig-13} for another plot of these data.
}
\end{figure}

\begin{figure}[p]
  \vbox to 12cm { \plotfiddle{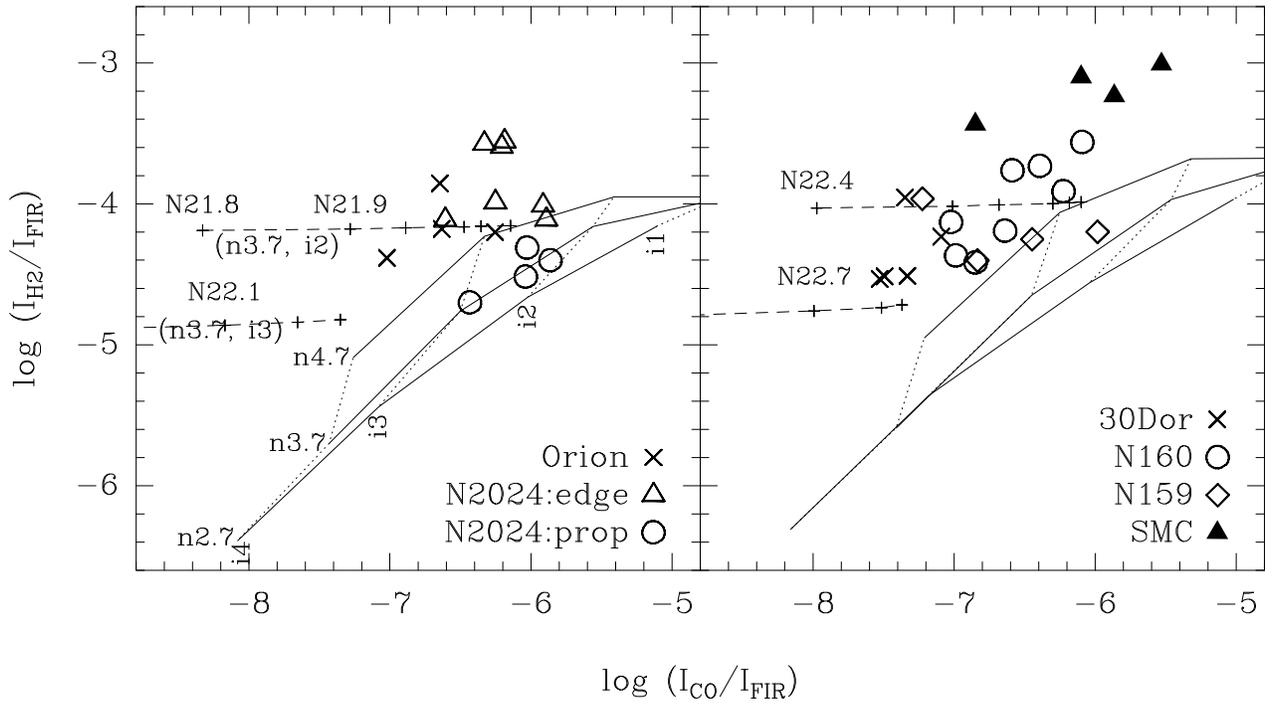} {12cm}{-90}{75}{75}{-285}{395} }
  \caption{ \label{fig-10}
    $\ihh/\ifir$ versus $\ico/\ifir$.
    See Figures \ref{fig-08} and \ref{fig-09} for the explanations.
  }
\end{figure}

\begin{figure}[p]
  \vbox to 13cm { \plotfiddle{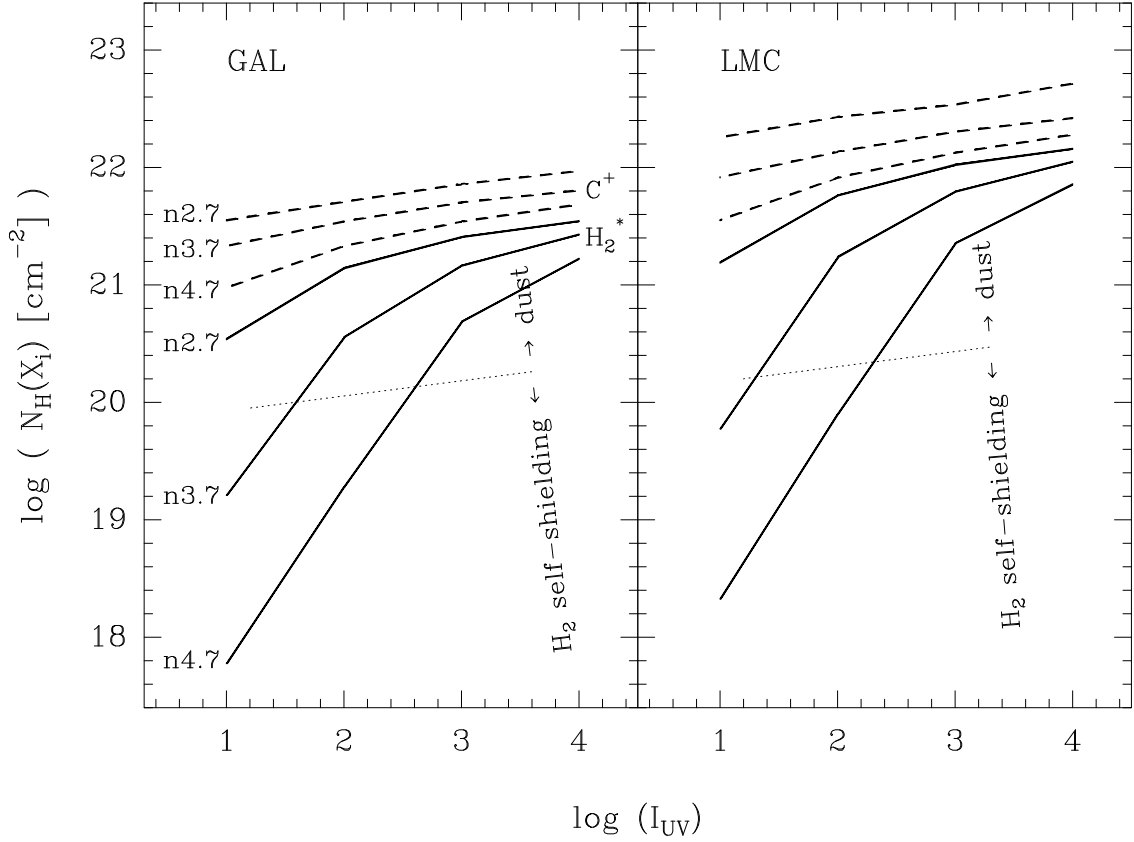} {13cm}{-90}{75}{75}{-285}{410} }
  \caption{ \label{fig-11}
    \iuv\ versus the depths ($N_H(X)$ = $X n_H$) from the surface of the cloud to the H-to-\hh\ transition layer (where the abundance of H becomes same as that of \hh, shown as solid lines), and to the \cplus-to-C-to-CO transition layers (where the abundance of \cplus\ becomes same as that of \twco, shown as dashed lines).
We changed the initial parameters, e.g., $n_H$, \iuv, and $Z$, for each model. $n2.7$ denotes $n_H$ $=$ $5 \times 10^2$ \cmv.
At fixed density, the dashed lines and the solid lines show the change of \xcplus\ and \xhhs\ respectively.
The dotted line divides the (\iuv, $n_H$, \xhhs) space into the \hh\ self-shielding dominant space and the dust absorption dominant space (see Equation~\ref{eq:self-shielding}).
}
\end{figure}

\begin{figure}[p]
  \vbox to 13cm { \plotfiddle{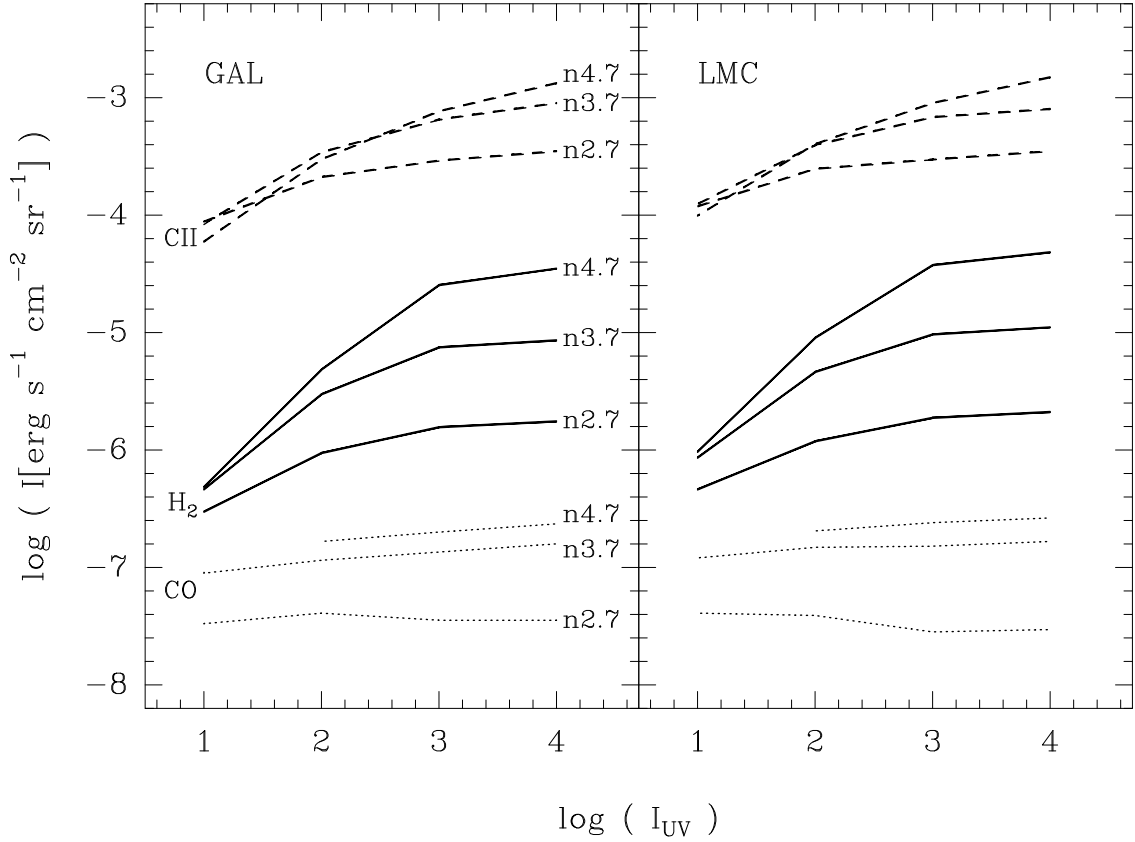} {13cm}{-90}{75}{75}{-285}{410} }
  \caption{ \label{fig-12}
\iuv\ versus [\cii] (dashed line), \hh\ \vone\ (solid line), and \twco\ \jone\ (dotted line) emission line intensities at fixed density.
We assume that $\eta = 1$ and the intensities do not include the geometry effect ($\epsilon_i$ = 1): $I_i$ = $f_i(n_H, \iuv, Z)$.
}
\end{figure}

\begin{figure}[p]
  \vbox to 14cm { \plotfiddle{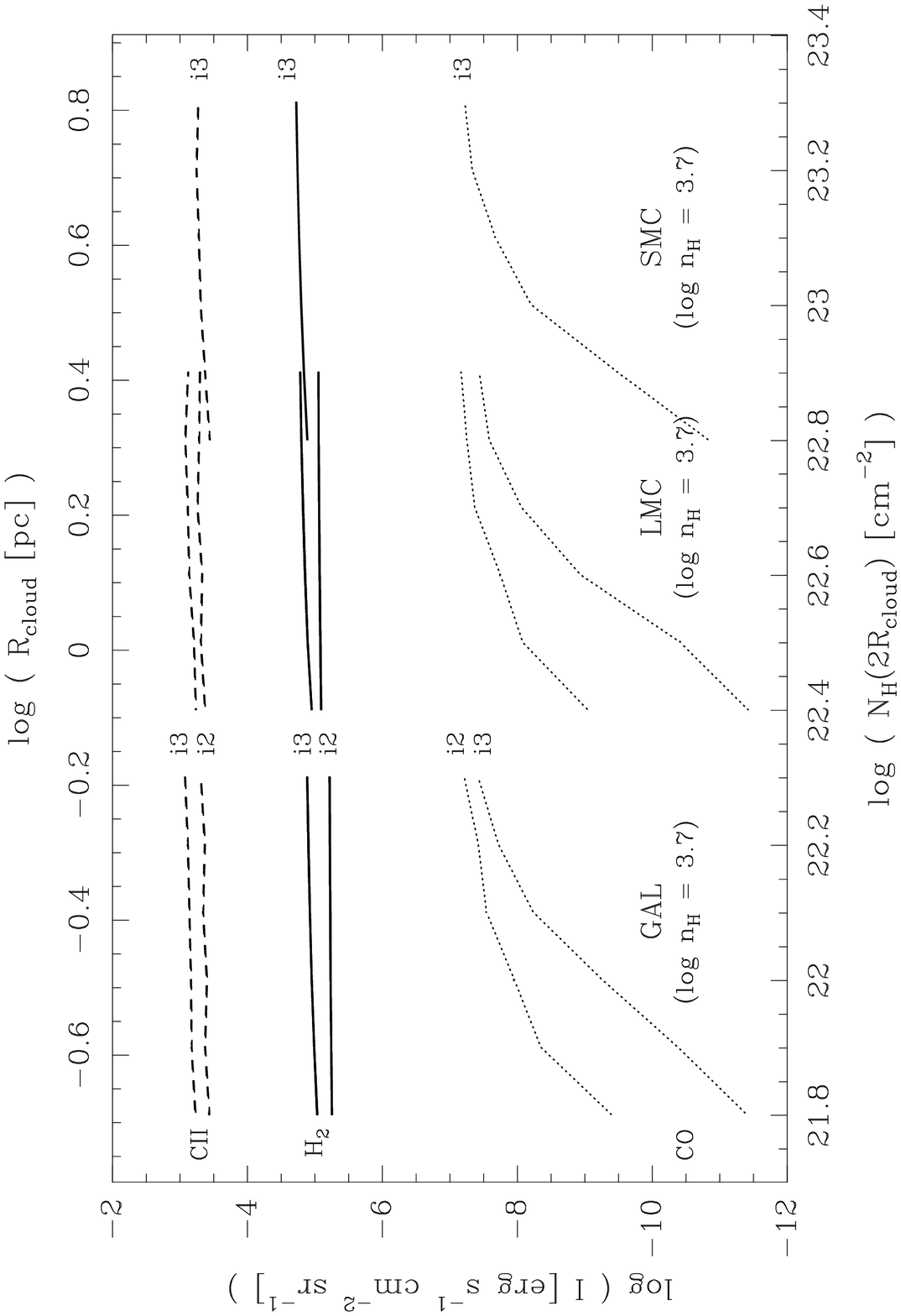} {14cm}{-90}{68}{68}{-255}{385} }
  \caption{ \label{fig-13}
[\cii] (dashed line), \hh\ \vone\ (solid line), and \twco\ \jone\ (dotted line) line intensities versus $R_{cloud}$ at ($n_H$ = $5 \times 10^3$ \cma; \iuv = $10^2$, $10^3$).
$I2$ and $i3$ denote \iuv\ = $10^2$ and $10^3$ respectively.
We assume that $\eta = 1$ and the intensities include the effect of spherical geometry: $I_i$ = $\epsilon_i f_i(n_H, \iuv, Z)$.
\icii\ and \ico\ are directly from the VT code, and \ihh\ is from $\epsilon_{H_2} f_{H_2}$ with Equations \ref{eq:f_H_2} and \ref{eq:eps1}.
We plot only the range of $N_H$ where significant changes take place in each model. 
The upper axis shows the cloud radius in pc, and the lower axis the central column density of the spherical cloud in \cma: $N_H(2R_{cloud})$ = $2 R_{cloud} n_H$.
Note that $N_H(X)$ in Figure \ref{fig-11} is $X n_H$, where \xcplus\ = $R_{cloud} - R_{CO}$.
}
\end{figure}

\begin{figure}[p]
  \vbox to 13cm { \plotfiddle{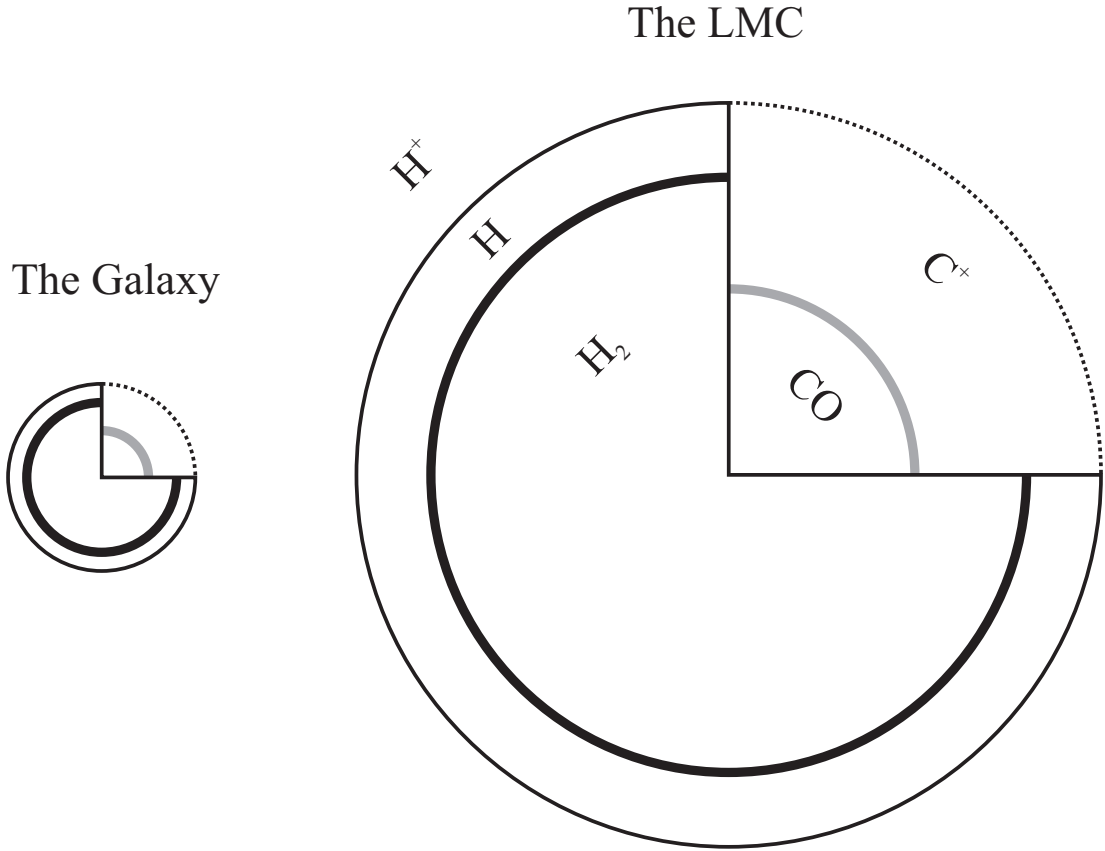} {13cm}{-90}{70}{70}{-310}{410} }
  \caption{ \label{fig-14}
   Schematic of chemical structures and sizes of typical clouds in the Galaxy and the LMC. Assuming that the far-UV field (\iuv) and the gas density ($n_H$) are same in the clouds in both galaxies, the cloud size in the LMC is four times bigger than that in the Galaxy, i.e., the typical size of the star-forming clouds is inversely proportional to the metallicity of the galaxy.
    }
\end{figure}

\begin{figure}[p]
  \vbox to 15cm { \plotfiddle{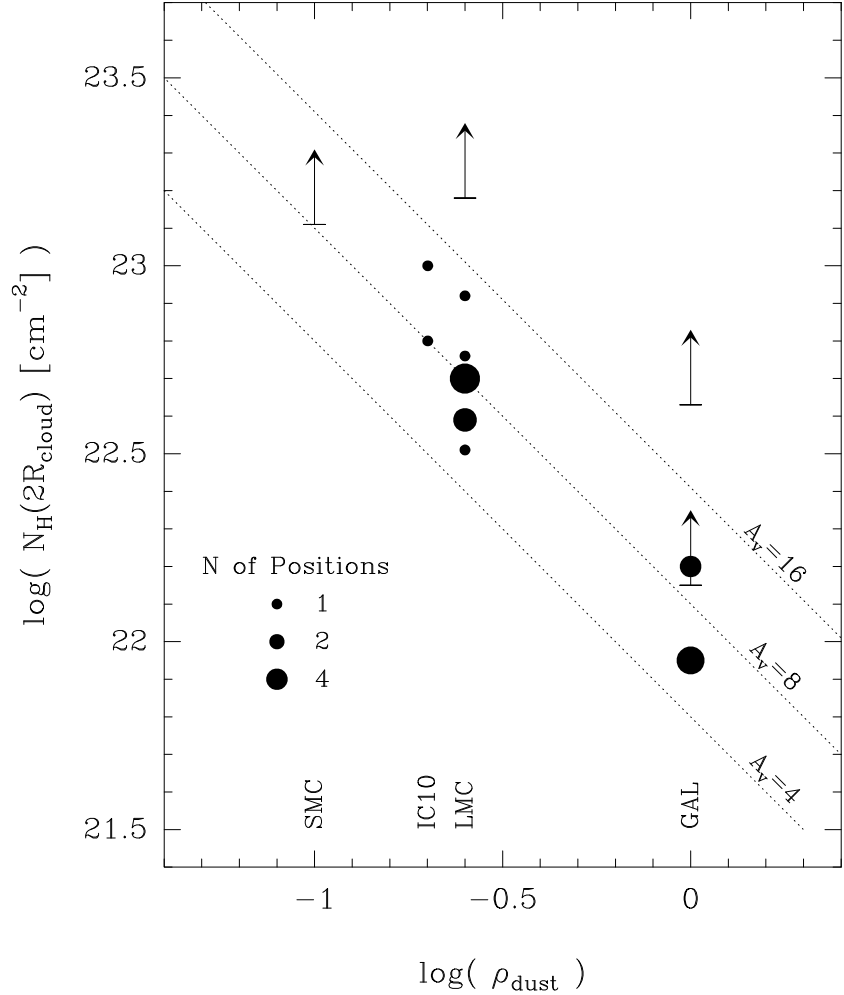} {15cm}{0}{72}{72}{-220}{-20} }
  \caption{ \label{fig-15}
    Central column density of spherical cloud versus dust-to-gas ratio from the Galaxy, the LMC, the SMC, and IC 10 (see also Table \ref{tbl-10}).
    Note that the cloud size is presented by the hydrogen column density for a given density, $n_H$ = $5 \times 10^3$ \cmv : $N_H(2R_{cloud})$ = $2 R_{cloud} n_H$, and $\rho_{dust}$ is normalized to the Galactic value. The area of the filled circle represents the number of observed positions. The arrow shows the lower limit. To compare with the theory of photoionization-regulated star formation, we overlay the constant optical depths at \av\ = 4, 8, and 16 mag.
  }
\end{figure}

\end{document}